\documentclass{aa} 
\usepackage{longtable}
\usepackage{latexsym}
\usepackage{amssymb}
\usepackage{lscape}
\usepackage[]{natbib}

\usepackage{graphicx}
\usepackage{txfonts}
\title{H$\alpha3$: an H$\alpha$ imaging survey of HI selected galaxies from ALFALFA\thanks{Based on observations 
taken at the  observatory of San Pedro Martir (Baja California, Mexico), belonging to the
Mexican Observatorio Astron\'omico Nacional.}}

\subtitle{II. Star formation properties of galaxies in the Virgo cluster and surroundings}

\author{G. Gavazzi \inst{1}
\and M. Fumagalli \inst{2,3}\thanks {Hubble Fellow}
\and M. Fossati \inst{4,1}
\and V. Galardo \inst{1}
\and F. Grossetti \inst{1}
\and A. Boselli \inst{5}
\and R. Giovanelli \inst{6}
\and M.P. Haynes \inst{6}
}

\authorrunning{G. Gavazzi et al.}
\titlerunning{H$\alpha3$: H$\alpha$ imaging survey of HI selected galaxies from ALFALFA}

\institute{Universit\`a degli Studi di Milano-Bicocca, Piazza della Scienza 3, 20126 Milano, Italy\\
\email {giuseppe.gavazzi@mib.infn.it}
\and
Carnegie Observatories, 813 Santa Barbara Street, Pasadena, CA 91101, USA\\
\email {mfumagalli@obs.carnegiescience.edu}
\and
Department of Astrophysics, Princeton University, Princeton, NJ 08544-1001, USA
\and
Max-Planck-Institut f{\"u}r Extraterrestrische Physik, Giessenbachstrasse, D-85748 Garching, Germany\\
\email {mfossati@mpe.mpg.de}
\and
Laboratoire d'Astrophysique de Marseille, UMR 6110 CNRS, 38 rue F. Joliot-Curie, F-13388, Marseille, France\\
\email {alessandro.boselli@oamp.fr}
\and
Center for Radiophysics and Space Research, Space Science Building, Ithaca, NY, 14853\\
\email {haynes@astro.cornell.edu, riccardo@astro.cornell.edu}
}

\begin{document}
\date{Received 8/1/2012; accepted 11/3/2013}

 
  \abstract
         {We present the analysis of H$\alpha3$, 
           an H$\alpha$ narrow-band imaging follow-up survey of 409 
           galaxies selected from the HI Arecibo Legacy Fast ALFA Survey (ALFALFA) 
           in the Local Supercluster, including the Virgo cluster,
	   in the region $\rm 11^h < R.A. <16^h\, ; \,4^o< Dec. <16^o$; $350<cz<2000$  $\rm ~km~s^{-1}$.}
	{Taking advantage of  H$\alpha3$, which provides the complete census 
         of the recent massive star formation rate (SFR) in HI-rich galaxies in the local Universe
	 and of ancillary optical data from SDSS
	 we explore the relations between the stellar mass, the HI mass, and the current, massive
	 SFR of nearby galaxies in the Virgo cluster. We compare these with those of isolated galaxies in the Local Supercluster,
         and we investigate the role of the environment in shaping the star formation properties of galaxies at the present cosmological epoch.}
         {By using the H$\alpha$ hydrogen recombination line as a tracer of 
           recent star formation, we  
           investigated the relationships between atomic neutral gas and newly formed stars
           in different environments (cluster and field), for many morphological types 
           (spirals and dwarfs), and over a wide range of stellar masses 
           ($10^{7.5}$ to $10^{11.5}$ $M_\odot$).
           To quantify the degree of environmental perturbation, we adopted an updated calibration of 
             the HI deficiency parameter  which we used to divide the sample into 
             three classes: unperturbed galaxies ($Def_{HI} \leq 0.3$), 
             perturbed galaxies ($0.3<Def_{HI}<0.9$), and highly perturbed galaxies ($Def_{HI} \geq 0.9$).}
         {
	  Studying the mean properties of late-type galaxies in the Local Supercluster, 
	  we find that galaxies in increasing dense local galaxy conditions (or decreasing projected angular
	  separation from  M87) show   
	  a significant decrease in the HI content and in the mean specific SFR, 
	  along with a progressive reddening of their stellar populations. 
	  The gradual quenching of the star formation occurs outside-in, consistently with the
	  predictions of the ram pressure model.   
    	  Once considered as a whole, the Virgo cluster is effective
    	  in removing neutral hydrogen from galaxies, and this perturbation is strong enough 
    	  to appreciably reduce the SFR of its entire galaxy population. }
         {An estimate of the present infall rate of 300-400 galaxies per Gyr in the Virgo cluster is obtained
	 from the number of existing HI-rich late-type systems, assuming 200-300 Myr as the time scale for HI ablation.
	 If the infall process has been acting at a constant rate, this would imply that the Virgo cluster has formed
	 approximately 2 Gyr ago, consistently with the idea that Virgo is in a young state of dynamical evolution.}        
   \maketitle
   \section{Introduction}

A deeper understanding of the basic processes that regulate
the life of galaxies, i.e, the conversion of gas into new stars and their progressive aging,
is becoming a reality owing to the recent multi-wavelength  
surveys that are providing a wealth of information on the different phases
of the interstellar medium (ISM), the stellar content, and the 
current star formation rate (SFR) in nearby galaxies. This observational
effort is complemented by new progress in numerical simulations and theory
in modeling the processes that regulate the formation of new stars in galaxies 
(e.g. Somerville \& Primack 1999).\\
ALFALFA (Giovanelli et al. 2005) is a blind survey that just ended at Arecibo (October 2012),
aimed at obtaining a census of HI sources within  7000 sq degrees of the sky accessible 
from Arecibo (including the Virgo cluster), with a sensitivity of
about 2\footnote{2 mJy is the noise per channel after Hanning smoothing to 10 $\rm km ~s^{-1}$
(see Haynes et al. 2011).} mJy $\rm km~s^{-1}$, corresponding to $10^{7.7} M_\odot$ of HI at the distance of Virgo.
A catalog listing 40\% of the whole ALFALFA sources ($\alpha.40$ catalog) has been published by Haynes et al. (2011).
The Sloan Digital Sky Survey (SDSS, York et al. 2000) revolutionized our knowledge on the stellar content  
of galaxies in the local Universe.
The GALEX mission (Martin et al. 2005) with its all-sky survey (and other smaller but deeper surveys,
such as GUVICS by Boselli et al. 2011) disclosed our view of the ultraviolet sky, thereby 
providing quantitative knowledge on the massive, recent  star formation in galaxies.\\     
\begin{figure*}[!t]
\centering
\includegraphics[width=19cm]{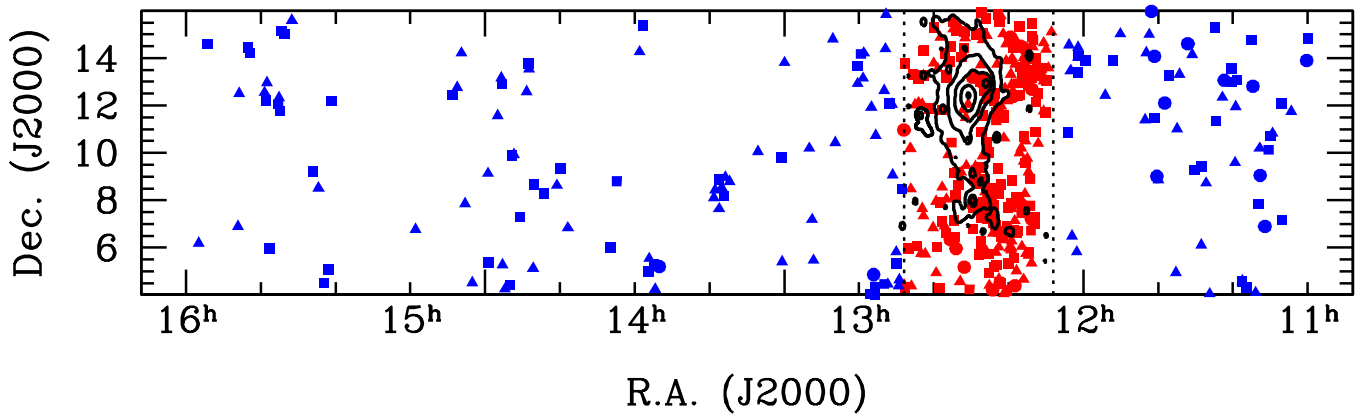}
 \caption{Sky distribution of 409 HI-selected galaxies in H$\alpha3$, coded according to their
  morphology (E-S0-S0a: circles; giant spirals Sa-Sm: squares; Irr-BCD: triangles)
The two vertical dashed	 lines mark the adopted boundaries of the Virgo cluster. 
Red symbols refer to galaxies considered as part of the Virgo cluster, blue are galaxies considered
as isolated. Superposed are the X-ray contours from ROSAT (B{\"o}hringer et al. 1994).}
\label{campione}
\end{figure*}
Based on these datasets Huang et al. (2012) were able to determine the scaling relations between
the atomic hydrogen content, the stellar content, and the SFR in HI selected galaxies
over a range of mass approaching one million (from $\rm 10^6 ~to~ 10^{12} M_{\odot}$).
Similar analyses from optically selected surveys were carryed out in the local volume 
(Lee et al. 2007, 2009, Bothwell et al. 2009) and beyond
(e.g. GASS, the GALEX Arecibo SDSS Survey, Catinella et al. 2010; Schiminovich et al. 2010),
limited  to galaxies with stellar masses in excess of $10^{10}$ $M_\odot$ however.\\
Huang et al. (2012) based their study on the star formation rate estimated by fitting 
spectral energy distributions (SEDs) obtained by
combining optical (SDSS) and UV (GALEX) photometry. This technique requires that extinction 
in the UV is properly 
accounted for (Cortese et al. 2006, 2008b) and depends on the assumed star formation history (SFH) and metallicity. 
On the other hand the UV or H$\alpha$ luminosity alone (see Kennicutt 1998)
provides a more istantaneous estimate of the SFR and
requires that the SFH of galaxies has remained stationary over the time scale
typical of A ($10^8$ yr) or OB stars ($10^{7}$ yr). 
This approach is more suitable when it is necessary to determine the SFR of galaxies 
that are subject to abrupt changes in their SFH, e.g., when they suffer from
environmental disturbances, such as quenching by ram pressure. 
Vollmer et al. (2004) and Boselli et al. (2006) have shown that in massive galaxies,
such as NGC 4569, ram pressure can produce complete quenching of the star formation on a timespan 
as short as 100-300 Myrs, and even shorter in dwarf galaxies (Boselli et al. 2008).\\
The SFR determined from UV magnitudes has been compared to that from recombination lines  by
Boselli et al. (2009), who arrived at the conclusion that the accuracy of SFR based on UV (GALEX) luminosity critically depends
on the quality of the data used, i.e, whether they come from the shallow (100 sec exposure)
All-sky-survey (AIS) or from the deeper MIS survey (typically 1500 sec integration).   
Huang et al. (2012) reported that 25\% of the $\alpha$.4 targets do not have a counterpart in GALEX,
52.2\% of the remaining match to UV sources found in the AIS, and only 23.1\% matches those in the MIS.
This severely hampers an unbiased comparison of the star formation properties of galaxies in the Virgo
cluster with respect to isolated galaxies in the Local Supercluster, i.e, the main 
goal of the present investigation using UV data, because the large majority of Virgo galaxies
have been observed in the MIS (owing to GUViCS by Boselli et al 2011), whereas the remaining objects
in the local field have only AIS data.
One more source of uncertainty in determining the SFR using SED fitting is connected with 
the large errors introduced by the SDSS pipeline on the photometry of large nearby galaxies due to 
``shredding" in multiple pieces, which leads to wrong magnitude determinations.\\
Given these considerations, we decided to make a new estimate of the "instantaneous" SFR based on 
our own measurements of the hydrogen recombination lines
using the data  from the H$\alpha3$ narrow-band imaging survey of nearby galaxies 
selected from ALFALFA in the spring sky of the Local Supercluster (Gavazzi et al. 2012a, Paper I), including the Virgo cluster.

Virgo is the cluster of galaxies nearest to us, which, in spite of the enormous angular 
extent of about 6x10 square degrees, 
 lies in the footprint of both ALFALFA and H$\alpha3$.
Already in early studies Virgo appeared to be a cluster in the early stage of dynamical evolution. 
Its general appearance is highly irregular 
(Abell et al. 1989), and it is classified as Bautz-Morgan type III (Bautz \& Morgan 1970). 
Owing to the intensive optical survey by Binggeli et al. (1985), its
velocity dispersion was found to be $760 \pm 45 ~\rm km ~s^{-1}$, lower than that of evolved clusters, 
indicating a shallow gravitational potential. The spiral fraction  
is 48\%  compared to 16\% of Coma. 
As seen by ROSAT (B{\"o}hringer et al. 1994), the Virgo cluster is known to have an irregular			 
X-ray morphology. 
These features provide evidence that the Virgo cluster is currently undergoing  dynamical evolution 
(Forman \& Jones 1982).  

The role of the environment in shaping the star 
formation properties of galaxies in the local Universe is still an open question. 
As early as in the Las Campanas redshift survey 
(Hashimoto et al. 1998), or in the DR3 release of the SDSS (G{\'o}mez et al. 2003), evidence has built
that suppression of the SFR of galaxies in dense environments takes place at $z$=0, up to $z$=0.8 (Patel et al. 2009) 
(see a review by Boselli \& Gavazzi 2006). 
By comparing the statistical HI  properties of galaxies 
in nine nearby clusters with those of "field" objects of similar type and luminosity, 
Giovanelli \& Haynes (1985) showed that clusters contain a high percentage 
of HI  deficient galaxies. 
The mechanism that most likely 
contributes to significant gas depletion in clusters is ram-pressure stripping (Gunn \& Gott 1972). 
Since then the HI  deficiency
parameter has been used as a proxy for the degree of the perturbation that galaxies
experience in the harsh environment of rich galaxy clusters.

Taking advantage of the presence of the Virgo cluster in H$\alpha3$,
the first goal of the present work is to quantify, in addition to the HI content, how the SFR
is affected by the environment of an evolving cluster such as Virgo, following the line traced by the
pioneering work of Kennicutt (1983). We approach this question by studying 
a possible residual correlation between the SFR, the gas content and the gas deficiency 
after the first-order scaling law from the stellar mass have been removed.   

The dataset used in this paper has been presented in
Paper I of the present series, which illustrates H$\alpha3$, the ongoing imaging survey
at the 2.1m telescope of the San Pedro Martir (SPM) Observatory,  
the details of the sample selection,						        
completeness, the H$\alpha$ observations, data reduction and analysis, which gives							        
access to the integrated extinction-free measurements of the SFR for 233							        
galaxies observed at SPM between 2006 and 2009. 
The present Paper II contains the analysis of the integrated SFR obtained using H$\alpha3$.
After a brief introduction to the galaxy sample (Section \ref{sample}),
we derive an updated calibration of the HI deficiency parameter valid for 
dwarf gas-rich systems such as those that dominate H$\alpha3$ (Section \ref{hidef}).
The main results of the present analysis are given in Section \ref{results},
where we analyze the relations between the mean optical colors and the extent of the star formation regions of
galaxies in the present survey as a function of their HI content. 
The scaling relations between gas and stellar masses and current SFR in the Local Supercluster are analyzed 
(Section \ref{scaling}).
The overall perturbation on the gas content and star formation properties of galaxies in the Local Supercluster due to the Virgo cluster 
is quantified in Section \ref{result}.
The results are discussed in Section \ref{discussion}, 
where some conclusions on the present evolutionary stage of the Virgo cluster are drawn, and 
we summarize our results in Section \ref{concl}.\\
Two companion papers continue this series. 
Paper III (Gavazzi et al. 2012b) carries an analysis similar to the one reported in the present paper, focused on 		        
the properties of galaxies in the Great Wall, including the Coma cluster.  							        
Paper IV (Fossati et al. 2012) contains the analysis of the structural properties of galaxies in the local and Coma Superclusters. 
The CAS parameters (concentration, asymmetry, and clumpiness developed by Conselice 2003) are determined on both the $r$ band and the   
H$\alpha$ images to study the morphology of the star-forming regions and compare it with that of older stars  			        
at arcsec resolution (unachievable on GALEX UV data which have a resolution of 4-5 arcsec).					        

Throughout the paper we adopt a flat $\Lambda CDM$ cosmology, with  
$H_0=73 \rm ~km~s^{-1}~Mpc^{-1}$ and $\Omega_\Lambda=0.7$.
All magnitudes are given in the AB system unless explicitly noted.

\section{Sample}
\label{sample}

\begin{figure*}[!t]
\centering
\includegraphics[width=18cm]{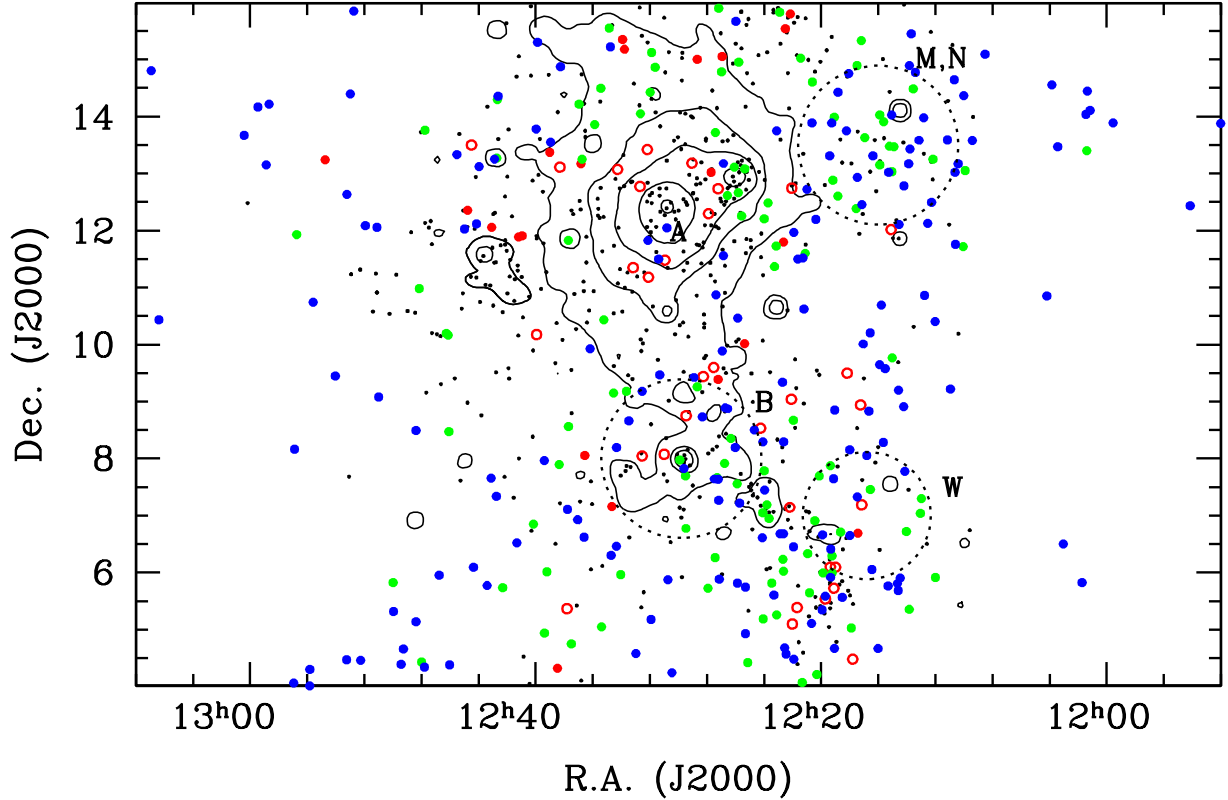}
 \caption{Sky distribution of galaxies around the Virgo cluster. ETGs (from the VCC with $m_p<$18.0 mag) are given with small black dots. 
 HI-selected LTG galaxies, coded according to their   HI deficiency parameter
 (red dots:  $Def_{HI}>0.9$; green dots: $0.3<Def_{HI}<0.9$; blue dots $Def_{HI} \leq 0.3$). The empty red
 circles represent the 26 LTGs undetected by ALFALFA listed in Table \ref{def}.
Superposed are the X-ray contours from ROSAT (B{\"o}hringer et al. 1994). The position of cluster A (M87), B (M49) 
and of clouds M, N and W are given. Clouds M and N coincide in position but have significantly different distance moduli.}
\label{virgo}
\end{figure*}

The sample analyzed in this work is drawn from the 900 square degree region 
$\rm 11^h <R.A. <16^h\, ; \,4^o< Dec. <16^o$, covering the Local Supercluster, including the Virgo cluster. 
This region has been fully mapped by ALFALFA  which provides us with a complete sample 
of HI-selected galaxies with gas masses as low as $10^{7.7} ~M_\odot$ (Haynes et al. 2011).

H$\alpha3$ is a follow-up survey consisting of 
H$\alpha$ imaging observations of ALFALFA targets with high signal-to-noise (typically S/N$~>6.5$)
and a good match between two independent polarizations (code = 1 sources; Giovanelli et al. 2005, 
Haynes et al. 2011). H$\alpha$ observations of members of the Virgo cluster were given in  
Gavazzi et al. (2002a, b, 2006), Boselli \& Gavazzi  (2002), and
Boselli et al. (2002b). Images and fluxes are also
publicly available via the GOLDMine web server (Gavazzi et al. 2003).
H$\alpha$ data for galaxies in low-density regions of the Local Supercluster are given in Paper I.
Outside Virgo the redshift window of H$\alpha3$ is $ 350<cz < 2000 \rm ~km~s^{-1}$, and
we limited the study to objects with HI fluxes $F_{\rm HI} > 0.7 {\rm~Jy~km~s^{-1}}$,
while in the Virgo cluster the velocity interval 
is extended to $ 350<cz < 3000 \rm ~km~s^{-1}$ to map the cluster in its full extent
we did not restrict ourselves to $F_{\rm HI} > 0.7 {\rm~Jy~km~s^{-1}}$.
At the distance of 17 Mpc, assumed for the Virgo subcluster A (Gavazzi et al. 1999), a flux limit  
$S_{21}=0.7 {\rm~Jy~km~s^{-1}}$ corresponds to an HI mass $M_{\rm HI}=10^{7.7} ~{M_\odot}$, 
while H$\alpha3$ is complete to $M_{HI}>10^{8}$ M$_\odot$ (see Paper I). Figure \ref{campione} 
illustrates the sky region covered by H$\alpha3$, which contains 409 galaxies. 

The basic quantities extensively used in this paper, i.e, the HI mass, the stellar mass, and the  
extinction-free SFR were derived in Paper I. 
The HI mass is defined by 
$M_{HI}= 2.36 \cdot 10^5 \cdot  S_{21} \cdot D^2$, 
where $D$ is the distance to the source in Mpc and $S_{21}$  is the integrated flux of the HI 
profile in units of Jy km s$^{-1}$ from ALFALFA.

As explained in Paper I, the $g$ and $i$ band total magnitudes of galaxies detected by ALFALFA
were measured by us on SDSS images (DR7, Abazajian et al. 2009) using IRAF.
This ensured proper foreground star-subtraction and overcomes the "shredding" 
problem generated by the automatic SDSS pipeline.

Inside the Virgo cluster we obtained an optically selected sample (see Section \ref{results}) of 640 objects based on the VCC catalog 
by Binggeli et al. (1985) 
(including objects of all morphological types with $B<18$ mag between $4^o<Dec<16^o$ and 
excluding those detected by ALFALFA
and galaxies in the background of Virgo with $cz>$3000 $\rm km~s^{-1}$). 
For this large sample  we computed the integrated magnitudes using an automatic procedure.
Starting from the SDSS images that are available from the GOLDMine web site, we produced deep ``white'' images by stacking  frames 
in five filters. Using SExtractor (Bertin \& Arnouts 1996), we separated the galaxies of interest from foreground stars and background galaxies.
After masking these objects, we computed the integrated magnitudes using elliptical apertures, the parameters of which
were computed on the white images. As a test, we compared the photometry derived with this automatic procedure with a subset of galaxies that
were measured manually on $i$ and $g$ images (Paper I) and we  obtained very consistent results. 

$g - i$ colors were corrected for internal extinction
using the empirical transformation dependent on stellar mass and 
galaxy inclination, derived in Appendix A of Paper III:
\begin{equation}
(g-i)_o = (g-i) - \Bigl\{ +0.17 \cdot [1-\cos(incl)] 
\cdot \Bigl[\log\Bigl(\frac{M_*}{M_\odot}\Bigr)-8.19\Bigr]\Bigr\}, \
\end{equation}
where $incl$ is the galaxy inclination (computed for disk galaxies (from Sa to Sdm)
following Solanes at al. (1996).

The stellar mass was derived from the $i$ magnitudes and $(g-i)_o$ color using the
transformation 

\begin{equation}
\log \Bigl(\frac{M_*}{\rm M_{\odot}} \Bigr) = -1.94 + 0.59 \cdot (g-i)_o + 
1.15 \cdot \log \Bigl(\frac{L_i}{\rm L_{\odot}} \Bigr)\
\label{eq:our_mass}
\end{equation} 
consistent with the mass determination of MPA-JHU\footnote{www.mpa-garching.mpg.de/SDSS/DR7/, see Salim et al. (2007)},
where $\log L_i$ is the $i$ band luminosity in solar units ($\log L_i=(I-4.56)/-2.5$).

The extinction-free SFR was computed from the luminosity of the H$\alpha$ line after correcting for 
Galactic and internal dust extinction and deblending from [NII] emission. 
Details on these quantities are given in Paper I.

Following a similar procedure as in Gavazzi et al. (2010, 2011) and Paper III, the local surface density 
$\Sigma$ around each galaxy was computed within a circle of 1 $\rm Mpc$ radius.
Around each galaxy we computed the density contrast as 
$$\delta_{1} = \frac{\Sigma-<\Sigma>}{<\Sigma>}$$,
where $\Sigma$ is the local surface density and $<\Sigma>$ = 1.3 gal $(\rm Mpc)^{-2}$ 
represents the mean surface density measured in the whole region. 

\section{HI content of Virgo galaxies}
\label{hidef}

 \begin{figure}[!t]
 \centering
  \includegraphics[width=8cm,height=8cm]{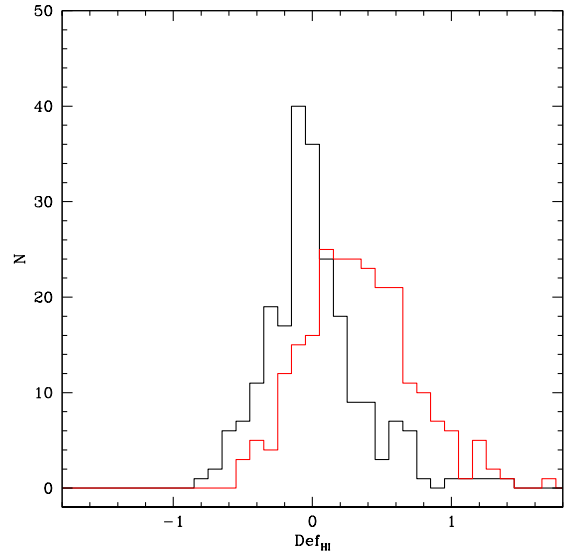}
\caption{Histogram of the $Def_{HI}$ parameter separately 
  for the isolated and W, N clouds (black) and for members of the Virgo A and B subclusters (red).
  The bulk of HI-deficient galaxies is located within the two subclusters.}
 \label{defhist}
 \end{figure}

The Virgo cluster is well known for its unrelaxed structure (Forman \& Jones 1982) and 
for its irregular morphology, elongated in the N-S direction. 
Figure \ref{virgo} shows an enlargement of Figure \ref{campione}, highlighting the zone that contains
the Virgo cluster, with the early-type galaxies (hereafter ETGs) added to emphasize the global density of galaxies in this region.
Following Gavazzi et al. (1999), we assume that Virgo  is composed of the main cluster A, which contains the bulk of the X-ray 
emission centered on M87, and
a secondary cluster B centered on M49, infalling into A at about 750 $\rm km ~s^{-1}$. These subcomponents
are at a distance of about 17 Mpc. Clouds M and W are twice more distant 
and are probably infalling into the main cluster along with cloud N, which is in the process of merging with cluster A.
Virgo is a spiral-rich cluster. 
The Virgo Cluster Catalogue (VCC, Binggeli et al. 1985) lists 48\%  (36\%) galaxies of type later than S0a for
$m_p<18$ mag ($m_p<20$). Restricting this to $m_p<13.9$ mag,  corresponding to the completeness limit 
of the spectroscopic SDSS $g<17.75$  at Coma, the spiral fraction increases to 60\%, i.e, four times more than
in the Coma cluster (16\%) at similar depth. 

It might well be that some of the late-type galaxies (hereafter LTGs) that contribute to the large spiral fraction in Virgo are 
LTGs seen projected on Virgo, but that are distributed along the line of sight on a filament connecting the 
local group to Virgo,
similar to some of the elongated filaments that exist in the Coma Supercluster (see Paper III). However, we estimate that
if we were to look at Coma from an angle such that we integrate along one of these filaments, the number of LTGs 
projected on Coma would increase from 16 \% to about 30\%. Although significant, this is not enough to explain the high 
fraction of LTGs seen in Virgo in terms of projection effects.

The parameter that is bestsuited to disentangle galaxies
that  physically belong to the Virgo cluster from either projected on it or belonging to 
clouds not yet fully processed by the cluster potential is perhaps the HI-deficiency parameter ($Def_{HI}$).
High-resolution HI maps of Virgo galaxies, such as those obtained by the VIVA survey of Chung et al. (2009),
confirm earlier claims (Cayatte et al. 1990) that 
the perturbed spiral/Irr galaxies in the Virgo cluster have a high $Def_{HI}$ and their HI 
morphology shows evidence of truncation/one-sidedness,
consistent with the hypothesis that they are suffering from ram-pressure stripping by the dense cluster ICM.

The $Def_{HI}$ parameter has been defined by Haynes \& Giovanelli (1984)
as the logarithmic difference between the HI mass observed in a galaxy 
and the expected value in isolated and unperturbed  objects of 
similar morphological type $T$ and linear diameter $d_L$:
$Def_{HI}=< \log M_{HI}(T,d_L)>- \log M_{HI}(obs)$. Here, 
$< \log M_{HI}(T,d_L)>= C_1+C_2\times 2 \log (d_L)$,
where $d_L$ (in kpc) is determined in the $g$ band at the $25^{th}$ $\rm mag ~arcsec^{-2}$ isophote.
The coefficients $C_1$ and $C_2$ were determined by 
Haynes \& Giovanelli (1984) by studying a control sample of isolated objects, 
and later, on a larger sample by  Solanes et al. (1996).   
Both samples  are composed almost exclusively of giant spirals however.
The HI-deficiency parameter  was therefore poorly calibrated for dwarf objects.

The problem is re-addressed here  using ALFALFA by assuming as isolated the galaxies 
outside the Virgo cluster (outside the interval 
$12^h08^m00^s<RA<12^h48^m00^s$), excluding a few more objects that appear to be clustered in small groups. 
We took these objects  (the blue symbols in 
Figure \ref{campione}) as a reference sample of nearly
isolated objects, representative of local normal and unperturbed galaxies. 
Scd-BCD dwarf systems dominate this sub-sample.\\ 
The coefficients $C_1$ and $C_2$ obtained by us for the giant galaxies (Sa-Sc) are similar to those of Solanes (1996) and of
Boselli \& Gavazzi (2009).
Consistent values are found in our sample for Scd-BCD. For this reason, 
we decided to adopt  
$C_1=7.51$ and $C_2=0.68$ for all LTGs (Sa-BCD), obtained by combining all late-type galaxies.
The resulting $Def_{HI}$ are plotted in 
Figure \ref{defhist}, where we display the frequency distribution of $Def_{HI}$ in the
isolated sample mixed with members of Virgo clouds M and W (black histogram), separately from Virgo members 
of sub-clusters A and B (red).
For the isolated and M, W clouds, we find a mean deficiency 
$<Def_{HI}>=0.01\pm 0.35$, consistent with normal HI content, while for sub-clusters A and B this is $<Def_{HI}>=0.33\pm 0.39$.

Unfortunately, the two distributions have overlapping tails, which reflecting  
the difficulty of spatially separating  in 3-D) galaxies that belong to the
real cluster from those that are just projected on it. For example (see Figure \ref{virgo})
there are many perfectly HI-normal LTGs projected within 2 degrees from M87, while there are several
highly deficient objects up to 4 degrees away from M87 or outside the X-ray contours. However,
the large majority of LTGs outside the X-ray contours are non-deficient (blue symbols).  
Thus the assumption that the HI-deficiency parameter is a reliable tracer of the degree of perturbation
suffered by LTG by the cluster medium, allows us to adopt  
in the rest of this paper a conservative threshold of $Def_{HI}$=0.3  (1 $\sigma$ above the
mean deficiency of the normal sample) to separate  normal from deficient galaxies.
There are approximately 30 galaxies in the bin $0.2<Def_{HI}<0.3$ that we consider normal, because they are
projected onto the N and W cloud. In reality they might belong to the outskirts of the A cluster,
which would justify their slightly positive deficiency.
Throughout this paper we adopt  $Def_{HI}=0.3$ as a threshold below which (270) galaxies can be considered as 
normal or unperturbed by the cluster environment, while those with $Def_{HI}> 0.3$
are treated as environmentally perturbed.
The abundant number statistics allows us to further sub-divide the deficient 
sample into $0.3<Def_{HI}<0.9$ (115 perturbed), and $Def_{HI} \geq 0.9$ (24 highly perturbed).

\section{Radio versus optical selection}
\label{results}

The comparison between the star formation properties of galaxies
selected with radio vs optical criteria
helps to shad some light on the relation between the star formation properties
(both global and nuclear) of LTGs and their HI content in various environments.

\begin{figure} 
 \centering
\includegraphics[width=9.5cm,height=9.5cm]{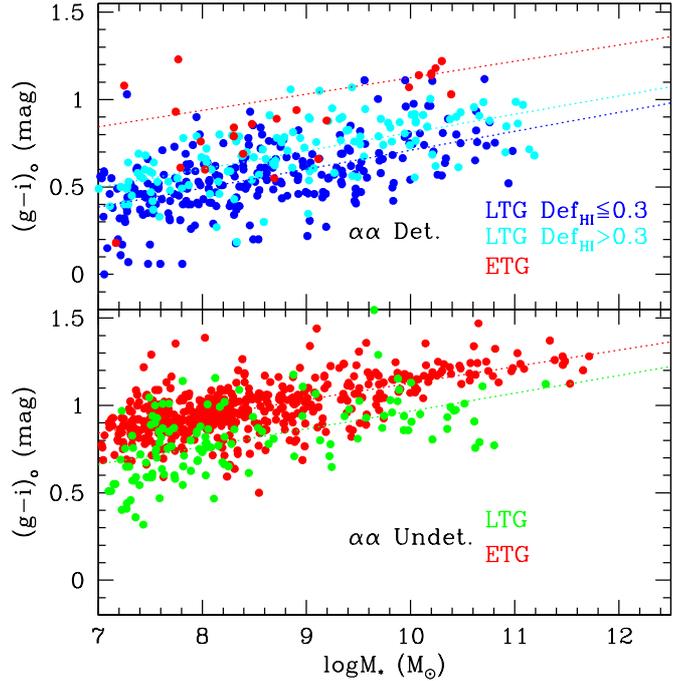}
\caption{
Top panel: ($g-i)_o$ color  (corrected for internal extinction) vs. stellar mass  diagram of H$\alpha3$ 
galaxies detected by 
ALFALFA (isolated+Virgo), color-coded by morphology (red=ETGs, blue+cyan=LTGs). 
Bottom panel: color-mass diagram of galaxies 
not in H$\alpha3$ (undetected by ALFALFA). They have been optically selected from NED among 
isolated objects, and from the VCC within the Virgo cluster (red = ETG; green=LTG). 
In both panels the diagonal dashed red lines mark the loci of the red sequence.
The 26 LTGs in the VCC, undetected by ALFALFA, with $logM_*>9.0$ are listed in Table \ref{def}.}
\label{colmag}
\end{figure}

As already stressed in Paper I, the radio selection biases H$\alpha3$  toward gas-rich, 
late-type galaxies (see also Gavazzi et al. 2008). 
The color ($g-i)_o$ versus stellar mass diagram composed
of galaxies in H$\alpha3$ (top panel of 
Figure \ref{colmag}) contains no well-developed red sequence
(as emphasized by the red dashed line 
that marks the loci potentially occupied by the red sequence).
This disagrees with similar diagrams derived from optically selected samples (e.g. Hogg et al. 2004). 
The census of H$\alpha3$ is made almost exclusively of LTG galaxies in the blue sequence. Only  
15 ETG galaxies are detected in the radio in our survey (see Figure \ref{colmag}, top panel).
Interestingly,  there is a minor (0.13 mag) but significant difference in the mean ($g-i)_o$  color between LTGs
with normal HI content (blue symbols in  Figure \ref{colmag}, top panel) and  LTGs with deficient HI content
(cyan symbols in  Figure \ref{colmag}, top panel).\\
 Furthermore, not all LTGs are detected by ALFALFA.
To demonstrate this point one would like to be able to extract a priori an optical catalog of galaxies in the
sky covered by ALFALFA.
A similar exercise was carried out by Cortese et al. (2008a) in their HI survey of the cluster A1367
and in Paper III for the Coma cluster.\\
Owing to the notorious limitations that affects the completeness of SDSS for $z<<0.05$ due to 
the shredding problem (Blanton et al. 2005), 
a similar exercise cannot easily be repeated for the Local Supercluster and Virgo.  
Alternatively, we proceeded in two steps as follows.
Out of the boundaries of the Virgo cluster we searched in NED all galaxies 
with $cz<2000$ $\rm km~s^{-1}$ in the footprint of H$\alpha3$. 
At their position we searched the SDSS navigator
and we took the $i$ and $g$ Petrosian magnitudes. 
After selecting those with $i<$17.0 mag, we were left with 50 objects, 35 of late and 15 of early type. 
It is not surprising that the 15
ETGs are not detected by ALFALFA. The 35 LTG have stellar mass  $logM_*<8.5$, thus it is not surprising
that they remain undetected by ALFALFA because, owing to the 
relation log$M_{HI}=0.55\times$ log$M_*+3.85$ that exists for unperturbed galaxies 
(see Section \ref{scale} and Gavazzi et al. 2008), they lie below the survey limiting sensitivity.\\
Inside the Virgo cluster a reliable optical selection was obtained using the VCC catalog by Binggeli et al. (1985) 
(including objects of all morphological type with $B<18$ mag (Vega) between $4^o<Dec<16^o$ and excluding those detected by ALFALFA
and galaxies in the background of Virgo  with $cz>$3000 $\rm km~s^{-1}$). 

For this large-sample we computed the integrated magnitudes using the automatic procedure described in Section \ref{sample}.
In the surveyed region we count 640 optically selected galaxies with $r<$17.5 mag (in addition to 
509 HI selected galaxies).
The color-mass relation of optical selected galaxies is plotted in the bottom panel of Figure \ref{colmag}
separately for early (red symbols) and late types (green symbols).
 The 26 massive LTG Virgo members (log$(M_*/M_\odot) > 9 $) undetected LTGs even though they are in the footprint of 
 ALFALFA\footnote{Six additional  VCC objects are undetected by ALFALFA, not because the are HI-poor, but
 because they are confused among each-other or with other sources: i.e, 
 VCC 386, 483, 497, 1673, 1676, and 1972; see Hallenbeck et al. (2012).}  are listed in Table \ref{def}.
 They are  well-known  high-HI-deficiency objects from previous deep pointed observations taken  at Arecibo by 
 Helou et al. (1984), Haynes \& Giovanelli (1986), Hoffmann et al. (1987,  1989), and Gavazzi et al. (2005).
 The interesting new fact is that the undetected LTGs have redder colors than the corresponding 
 HI-detected galaxies (see bottom panel of Figure \ref{colmag}), and are approximately 0.2 mag bluer than  
 galaxies in the red sequence, which fill the ``green valley'' (see also Cortese \& Hughes 2009).

All fits have a consistent slope of 0.10 and the intercepts are:  
0.18 for ETGs; -0.05 for undetected LTGs (0.23 mag bluer); -0.24 for detected LTGs with $Def_{HI}>0.3$  (0.19 mag bluer);
and -0.37 (0.13 mag bluer) for detected LTGs with $Def_{HI}<0.3$.
 A consistent increase of the specific star formation rate (SSFR) in steps of decreasing $Def_{HI}$ is 
also visible in Figure \ref{scale} (c).
\begin{figure}[!t]
\centering
\includegraphics[width=9.5cm]{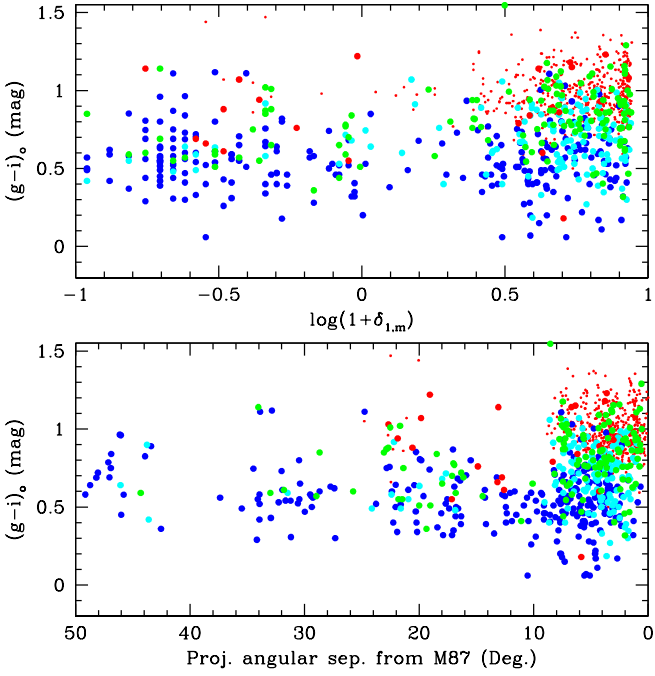}
 \caption{Corrected  ($g-i)_o$ color of galaxies of all morphological types as a  function of the 
 local galaxy density  (top panel) and
  of the projected  angular separation from M87 (bottom panel). 
 HI-rich LTGs are depicted with blue symbols; HI-poor LTGs are plotted with cyan symbols; LTGs undetected 
 by ALFALFA are given with green symbols;
 ETGs are marked with red symbols.}
\label{tutte}
\end{figure}
 The above exercise, although it should be taken with a grain of salt because of NED 
 incompleteness, confirms the result of Cortese et al. (2008a) (A1367) and of Paper III (Coma):
 given the sensitivity of ALFALFA at the distance of the Local Supercluster, only LTGs
 are detected. Among them, however, ALFALFA misses objects that are the most HI-anemic. \\
 Another way to show how the morphological/HI-content mix of galaxies depends on the environment in 
 and around the Virgo cluster is offered in
 Figure \ref{tutte}. Here  we plot the corrected color of galaxies of all morphological types
 regardless of their HI properties (i.e, including objects undetected by ALFALFA) as a function 
 of the projected separation from M87 
 (bottom panel) and of the local galaxy density (top panel). We use the same symbols as in Figure \ref{colmag}.
 The Virgo cluster (which lies within 10 degrees from M87) is visible 
 because it causes a strong color and morphology segregation,
 i.e, all ETGs (in the red sequence) are confined into its boundaries. Moreover, it also segregates LTGs with a low 
 HI content from their HI-rich counterparts, which are mostly found outside its boundaries.
 These families also appear to be significantly segregated as far as their specific star formation is concerned, 
 as indicated by their different color.
\begin{figure} 
 \centering
\includegraphics[width=9.5cm]{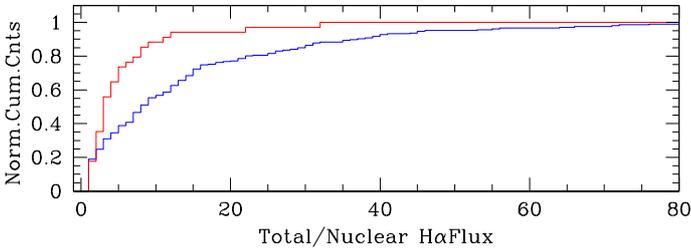}
\caption{
Cumulative distribution of the ratio of total H$\alpha$ flux in the imaging material to the H$\alpha$ 
flux measured in a central aperture
 of 10, 15, 30 or 45 arcsec diameter for galaxies of major axis  0-100, 100-200, 200-300 and
$>$ 300 arcsec.  
The blue histogram refers to LTG-in ALFALFA with $Def_{HI} \leq 0.3$ and the red one includes LTGs with $Def_{HI}>0.9$
and LTGs undetected by ALFALFA.
}
\label{fluxratio}
\end{figure} 
\begin{table}
\caption{26 LTG VCC galaxies in the Virgo cluster with log$M_*>9.0$ not detected by ALFALFA,
with detections or more stringent upper limits from previous pointed observations. A classification of the
nuclear spectrum is given in Column 6 and the morphology of the H$\alpha$ source in Column 7.
}
\begin{center}
\begin{tabular}{lrrrrrr}
\hline
VCC &  Type  &  log$M_*$  &    log$M_{HI}$  &  $Def_{HI}$ &	Nuc Sp   & H$\alpha$ morp. \\ 
\hline
(1) &  (2)   & (3)          &    (4)  &  (5)        &	(6)      &  (7)   \\  
\hline
341  &  Sa   &	10.41 &$<$  7.63    & $>$1.74 & RET  &     nuc+bar    \\    
358  &  Sa   &	 9.80 &$<$  8.18    & $>$0.77 & PAS  &     undet      \\    
492  &  Sa   &	10.00 &$<$  7.79    & $>$1.33 & LIN  &     circnuc    \\    
517  &  Sab  &	 9.09 &     7.40    &    1.17 & -    &     disk       \\    
522  &  Sa   &	 9.57 &$<$  7.44    & $>$1.62 & PSB  &     -          \\    
524  &  Sbc  &	10.35 &     8.15    &    1.42 & LIN  &     disk       \\    
534  &  Sa   &	 9.82 &     7.64    &    1.44 & AGN  &     circnuc    \\    
713  &  Sc   &	 9.97 &     8.10    &    1.38 & HII  &     pointl     \\    
984  &  Sa   &	 9.93 &$<$  7.31    & $>$1.82 & PAS  &     circnuc    \\    
1017 &  Im   &	 9.77 &$<$  7.10    & $>$2.22 & -    &     -          \\    
1047 &  Sa   &	10.00 &$<$  7.44    & $>$1.49 & PAS  &     -          \\    
1086 &  S..  &	 9.87 &     8.06    &    1.26 & AGN  &     disk       \\    
1158 &  Sa   &	10.35 &$<$  7.13    & $>$2.08 & PAS  &     -          \\    
1190 &  Sa   &	10.66 &$<$  7.64    & $>$1.83 & LIN  &     circnuc    \\    
1326 &  Sa   &	 9.49 &$<$  7.24    & $>$1.65 & HII  &     -          \\    
1330 &  Sa   &	 9.89 &     7.94    &    0.98 & PAS  &     disk       \\    
1412 &  Sa   &	10.43 &$<$  7.02    & $>$2.30 & PAS  &     disk       \\    
1419 &  S..  &	 9.24 &$<$  7.27    & $>$1.70 & HII  &     pointl     \\    
1435 &  Im   &	 9.22 &$<$  6.58    & $>$2.05 & -    &     -          \\    
1448 &  S?   &	 9.68 &$<$  7.45    & $>$1.31 & PAS  &     undet      \\    
1486 &  S..  &	 9.32 &     7.66    &    0.96 & HII  &     disk       \\    
1552 &  Sa   &	10.09 &$<$  7.16    & $>$2.15 & LIN  &     undet      \\    
1730 &  Sc   &	 9.93 &     7.83    &    1.11 & LIN  &     circnuc    \\    
1757 &  Sa   &	 9.36 &     7.38    &    1.51 & HII  &     pointl     \\    
1813 &  Sa   &  10.69 &$<$  7.19    & $>$2.18 & RET  &     undet      \\    
1999 &  Sa   &	 9.74 &$<$  7.19    & $>$1.74 & PAS  &     undet      \\    
\hline
\end{tabular}
\end{center}
\label{def}
\end{table}
\begin{table}
\caption{Frequency of nuclear activity in LTGs detected by ALFALFA with log$M_*>7.0$ 
(blue symbols in Fig \ref{colmag})
divided in three classes of increasing $Def_{HI}$. The undetected LTGs (green symbols in Fig \ref{colmag})
are combined with the $Def_{HI}$  $>$0.9. The last column gives the percentage of available nuclear spectra. 
}
\begin{center}
\begin{tabular}{lccccc}
\hline
$Def_{HI}$           &  \%  PASS  &  	 \% AGN+LIN & \%  HII	& compl.\%   \\
\hline
$<$0.3      	     &    7 	  &  	     8  &   85	    & 	 85   \\
0.3-0.9              &    14 	  &  	    13  &   73	    & 	 83   \\
$>$0.9               &    24 	  &  	    15  &   46	    & 	 78   \\  
\hline
\hline
\end{tabular}
\end{center}
\label{nuc}
\end{table}
The properties of the studied galaxies on nuclear / circumnuclear scales are investigated in 
Figure \ref{fluxratio} and in Table \ref{nuc}.
The figure gives the cumulative distribution of the ratio of total H$\alpha$ flux (measured in the imaging material) 
to the H$\alpha$ flux measured in a central aperture of variable size according to the total size of the galaxy,
separately for LTGs-in ALFALFA with $Def_{HI} \leq 0.3$ (blue) and for LTGs with $Def_{HI}>0.9$
mixed with LTGs undetected by ALFALFA (red). 
A Kolmogorov-Smirnov test indicates that the probability that the two samples are drawn from the same parent population
is 0.03 \%, i.e, that they significantly differ. In other words,  the  H$\alpha$ 
sources associated with HI-rich galaxies are extended, while those of strongly deficient/undetected objects 
are predominantly  nuclear/circumnuclear, see also Koopmann \& Kenney (2004a,b).\\
Table \ref{nuc} summarizes the classification of the available nuclear SDSS spectra of LTG galaxies 
in this work in three bins of
$Def_{HI}$ (the LTG galaxies undetected by  ALFALFA are included in the highest deficiency bin).
It shows that the fraction of star-forming nuclei decreases significantly with increasing $Def_{HI}$, 
although some signs of nuclear star
formation remain in 46 \% of the most HI-poor systems. The fraction of passive spectra is relevant 
only in the highest $Def_{HI}$ bin, while
the  fraction  of AGNs (including LINERS) does not significantly change with $Def_{HI}$ (consistently with Gavazzi et al. 2011).\\
Altogether we find evidence that\\
- there is a mild gradual progression toward redder colors from (1) LTGs detected by ALFALFA with normal HI content, 
(2) LTGs detected by ALFALFA with deficient HI content, (3) LTGs undetected by ALFALFA and (4) ETGs;\\ 
- the same morphology/HI-content sequence depends on the ambient density;\\
- the extent of the star formation region in HI-rich (1) LTGs is statistically larger than in their
HI-poor (2+3) counterparts.\\
In other words, we confirm the finding of Paper III (Coma) that the environmental conditions
not only affect the HI content of galaxies, but also govern the average color (specific star formation rate)
of galaxies. Moreover, whatever ablation mechanism is causing this pattern, it must proceed outside-in, 
in agreement with Boselli et al. (2008).

 \section {Scaling relations}
 \label{scaling}

The study of the scaling relations among the barionic constituents 
of (gas rich) galaxies has made a significant step forward with the work of Huang et al. (2012),
who analyzed $\alpha$.40 (Haynes et al. 2011), the most updated ALFALFA catalog  available so far.
Here we compare our local data with the results of Huang  et al. (2012),
whereas in Paper III we compare Virgo with Coma. 

\begin{figure} 
 \centering
\includegraphics[width=9.5cm]{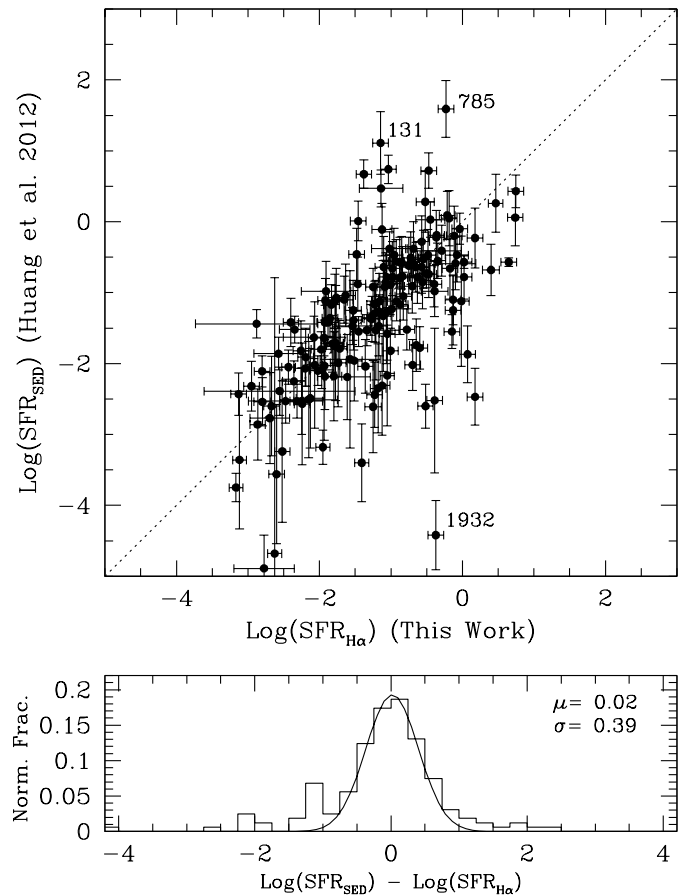}
\caption{(Top) Correlation between the SFR used in this work (from H$\alpha$) and those adopted by Huan et al. (2012)
(computed from SED fitting) 
The one-to-one relation is given with a dotted line.
Some outliers marked with their VCC names are commented on in the text. (Bottom) The distribution of the residual
of the two star formation indicators, with a Gaussian of mean $\mu$ and $\sigma$ fitted to it.
}
\label{huang}
\end{figure}

We begin by comparing in Figure \ref{huang} 
the star formation estimates obtained in the
present work from the H$\alpha$ corrected luminosities taken from Paper I,  
with the estimates of Huang et al. (2012)
computed by fitting UV, $u,g,r,i,z$ SEDs. 
The correlation cefficient is 0.65, 
implying that the two quantities have a probability higher than 99 \% of being correlated.
The bottom panel shows the distribution of the differences of the two quantities.
The Gaussian fitted to the data was obtained after sigma clipping of the most deviant points.
Two most extremes outlier galaxies (VCC 131 and VCC 1932) merit a more detailed comment.
These are both bright edge-on galaxies with significant H$\alpha$ flux (see GOLDMine) 
(logF=-12.68 and -12.32 $\rm erg~cm^{-2}s^{-1}$ respectively), therefore
with consistent SFR as computed in this work (-1.15 and -0.37 $\rm M_\odot yr^{-1}$). 
Furthermore the SFR derived by directly converting the FUV 
magnitudes (corrected following Cortese et al. 2006, 2008b, assuming balance between the absorbed UV photons and the total FIR energy)
into SFR are found to be (-0.44 and 0.31) within a factor of 5 from each other.  
Conversely, the SFR found by Huang et al.  (2012) 
(1.11, -4.42) differs by more than five orders of magnitude, revealing some problems that are probably connected to the SED fitting procedure.
Similarly, the galaxy VCC 785 is discrepant in Figure \ref{huang}, the SFR by SED fitting is approximately 
two orders of magnitude above the linear correlation. 
This is a face-on galaxy surrounded by a faint outer disk
(as revealed in a deep 1500 sec NUV image by GALEX not accompanied by a similar length FUV exposure). 
However, as explicitly mentioned by Huang et al. (2012), when deep GALEX images are not available in 
both NUV and FUV bands, the single long exposure is discarted and the SED fitting
is performed with available FUV and NUV data of similar exposures, albeit shorter. For VCC 785 the GALEX database reports
a FUV magnitude of 19.5 mag compared with a NUV of 15.4 mag (as derived from 200 sec exposures). The SED fitting routine probably 
assumes an overestimated extinction correction to allow for such a large discrepancy in magnitude, hence it overestimates the SFR.
In summary, we conclude that, albeit the large scatter, the two sets of SFR measurements are consistent,
which makes the comparison between the scaling relations obtained from the two SFR estimators meaningful.

\begin{figure*}
   \centering
   \includegraphics[width=6.0cm,height=6.0cm]{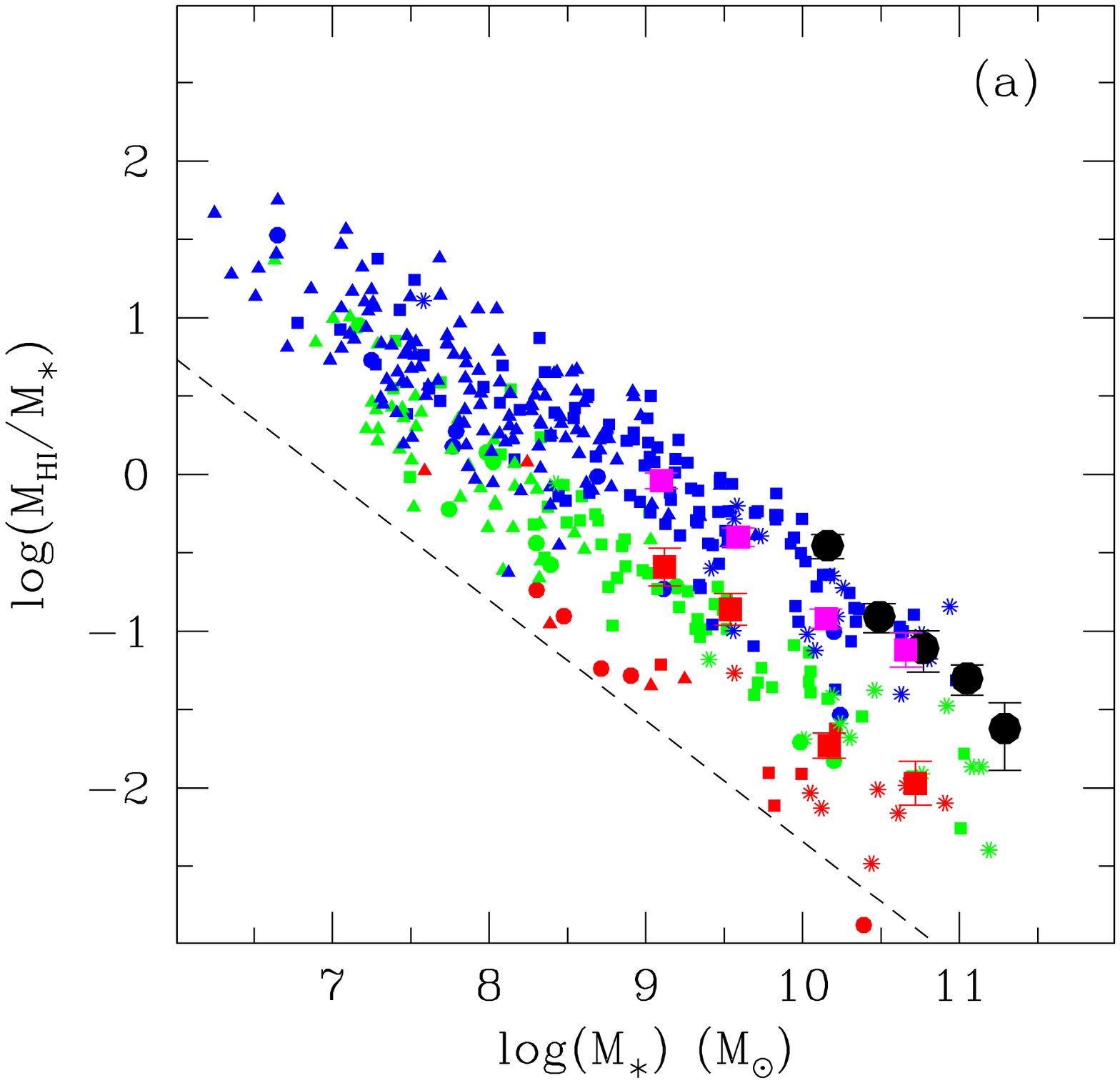}
    \includegraphics[width=6.0cm,height=6.0cm]{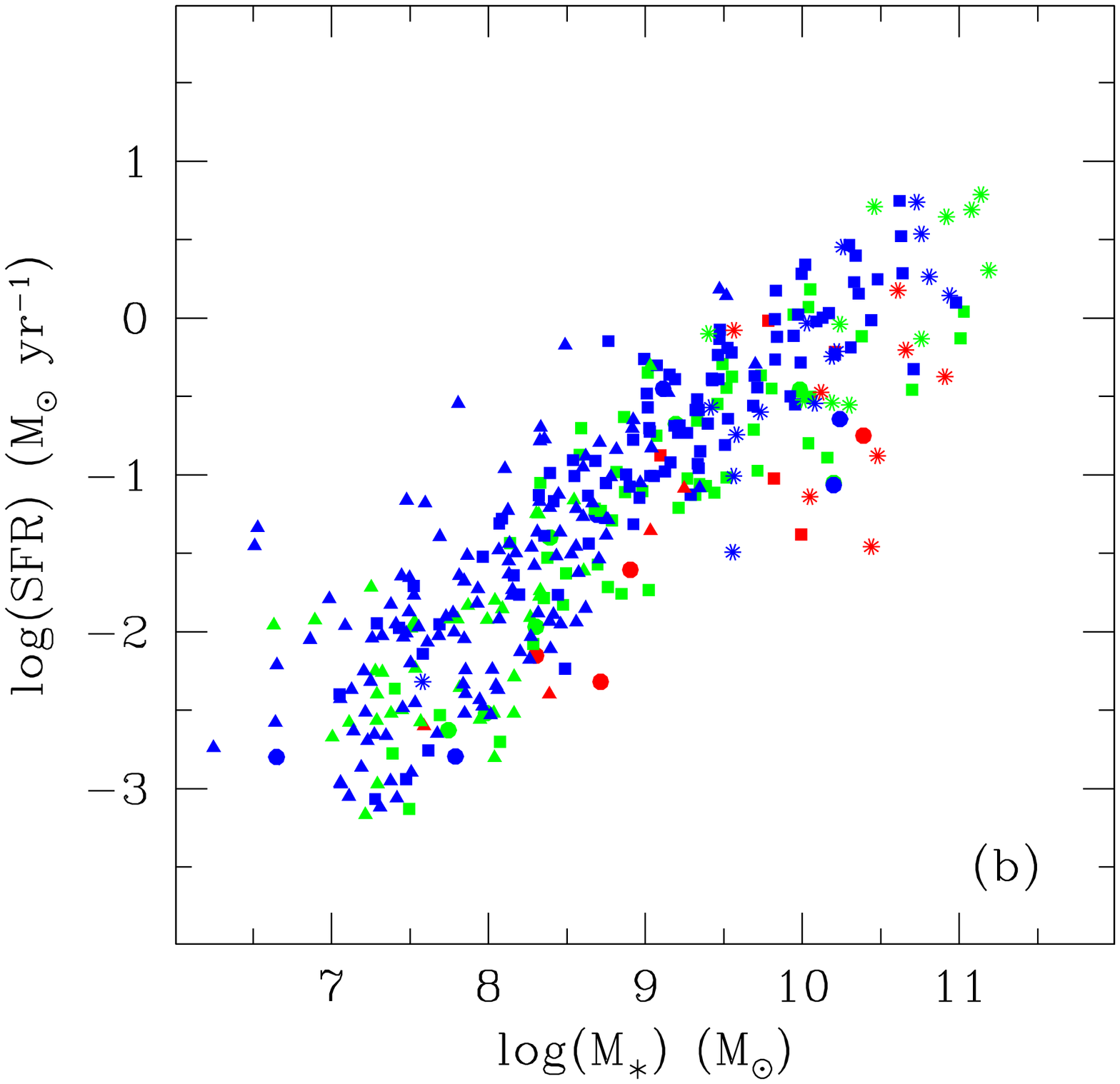}
    \includegraphics[width=6.0cm,height=6.0cm]{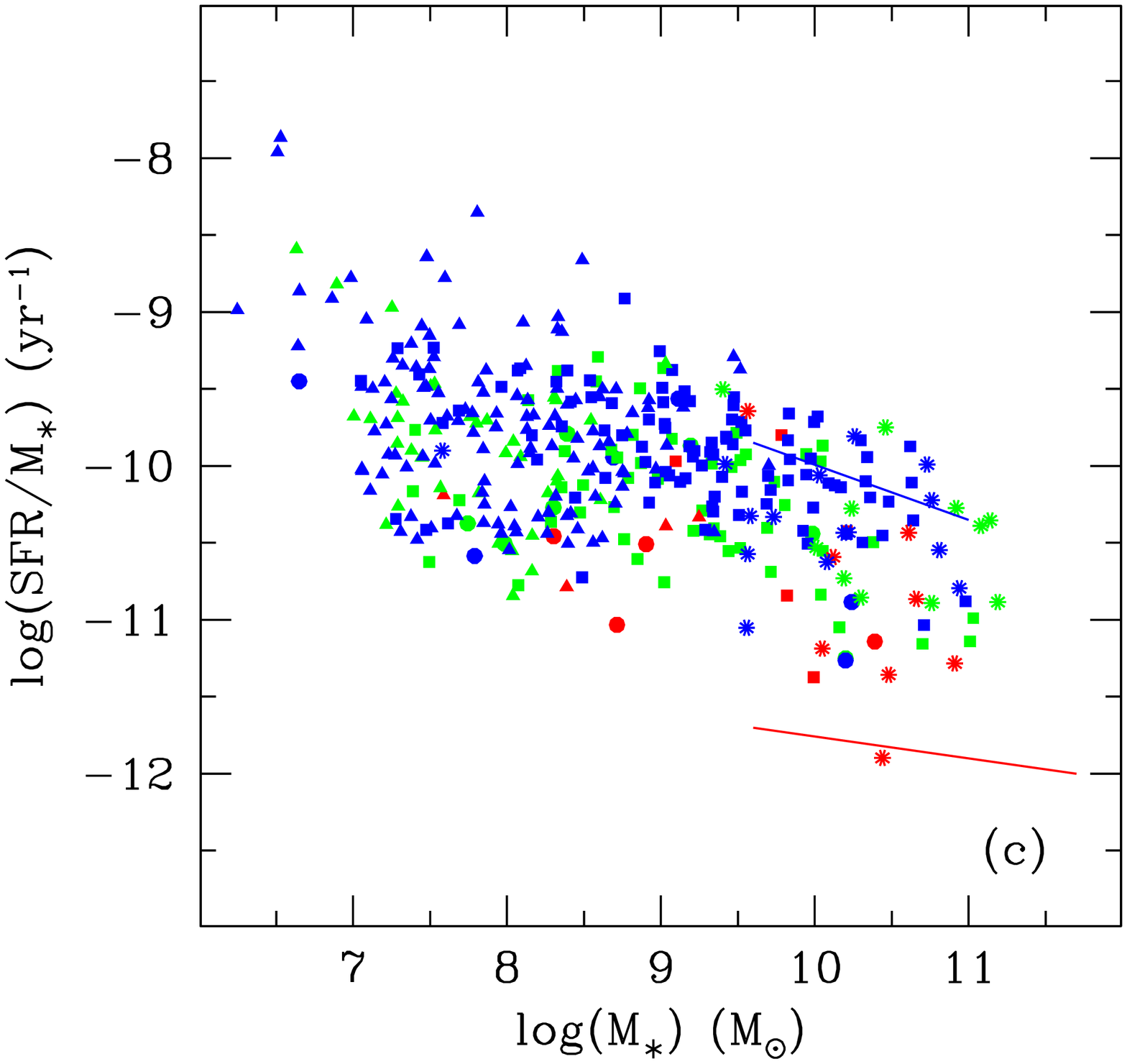}
    \caption{Scaling relations of HI mass, stellar mass and SFR   for galaxies in three classes 
    of increasing $Def_{HI}$. blue: $Def_{HI} \leq 0.3$ (normal); green: $0.3<Def_{HI}<0.9$ (perturbed), 
  red: $Def_{HI} \geq 0.9$ (highly perturbed).   Symbols are assigned according to the
  morphology. E-S0-S0a: circles; giant spirals (Sa-Sm): squares; Irr-BCD: triangles; AGN: asterisks. 
Panel (a): gas fraction ($M_{HI}$/$M_*$) versus $M_*$. 
    The diagonal dot-dashed line represents the limiting sensitivity of ALFALFA at the distance of Virgo.
    The black symbols give the $representative-sample$ extracted from GASS by Catinella et al. (2010), 
    the large magenta squares represent the HI-normal sample and the large red squares the Virgo 
    sample from Cortese et al. (2011).
    Across the stellar mass range in common there is an almost perfect agreement, but the present 
    sample extends the correlation to two orders of magnitude fainter.  
Panel (b):  SFR versus $M_*$. 
Panel (c):  Specific SFR versus $M_*$. 
The blue and red solid lines reproduce the relations plotted in Figure 8 of Schiminovich et al. (2010) 
for the star-forming sequence (blue) and the non-star forming locus of red galaxies (red).
}
\label{scale}
 \end{figure*}

 The scaling relations relevant to our study are given in Figure \ref{scale} (a) through (c), 
  and the linear regression parameters are listed in Table \ref{fit}\footnote{Linear regressions are obtained in 
 this work using the "bisector" method by Isobe et al. (1990) 
(mean coefficients of the direct and the inverse relation).} .

  We consider the relation between  the HI and stellar mass
  by showing in Figure \ref{scale} (a) the HI  gas fraction $M_{HI}$ /$M_*$ as 
  a function of the stellar mass.
  The diagonal dot-dashed line represents the sensitivity limit of ALFALFA, 
  computed for galaxies with an inclination of 45 deg in the plane of the sky
  \footnote{As introduced in Giovanelli et al. (2005),  ALFALFA is
  a noise-limited survey rather than a flux-limited one. At any given integrated HI  mass 
  the 21 cm flux per velocity channel is inversely proportional to the width of the HI  profile, hence
  to the galaxy inclination.  The completeness and sensitivity of ALFALFA are clearly defined and discussed
  in detail in   Saintonge (2007), Martin at al. (2010) and Haynes et al. (2011).}.
  By slicing the  H$\alpha3$ sample in three subsamples of increasing $Def_{HI}$:
 $Def_{HI} \leq 0.3$ (normal; blue symbols); $0.3<Def_{HI}<0.9$ (moderate deficiency; green symbols), and $Def_{HI} \geq 0.9$ 
 (high deficiency; red symbols)  we explore the whole range of environmental perturbations exhibited by galaxies in the 
 Local Supercluster, an aspect that was not examined by Huang et al (2012).\\
 Consistently with the results of Gavazzi et al. (2008) and Huang et al (2012),
 the gas to stars fraction decreases by approximately 4 orders of magnitude by increasing galaxy  mass, from log$M_*\sim 7$ M$_\odot$
 to $\sim 11.5$ M$_\odot$, independently of the HI deficiency parameter.
  This basic result seems to be independent of the fact that  H$\alpha3$ is an HI-selected survey. Indeed
 our scaling relation perfectly agrees with the one found in the GASS survey by Catinella et al. (2010)
 (see the black symbols in Figure \ref{scale}(a)).
 Similarly,  over a broader interval of stellar mass $10^{8.5}<M_{*}<10^{11.5}$ M$_\odot$ there is also a good agreement with 
 the  relation  found by Cortese et al. (2011) among HI-normal galaxies in
 a near-infrared selected sample  derived from the Herschel Reference Survey (HRS; Boselli et al. 2010) 
 (see the large magenta squares in Figure \ref{scale}(a))\footnote{The GASS and HRS points 
 plotted in Figure \ref{scale}(a) 
 are binned averages, thus the error-bars do not represent the dispersion of the distribution.}.
 This result is consistent with the $downsizing$ scenario (Gavazzi 1993, Gavazzi et al. 1996,
 Gavazzi \& Scodeggio 1996, Boselli et al. 2001,  Fontanot et al. 2009), where progressively more massive 
 galaxies have less gas/star at $z$=0 because they have transformed most of their gas into stars at higher $z$, 
 while late-type dwarf galaxies retain large quantities of hydrogen capable of 
 sustaining the star formation at some significant rate at the present cosmological epoch.

 The log$SFR$ versus log$M_*$ relation  (Figure \ref{scale}(b)) obtained at $z$=0 is near the direct proportionality 
 (slope=0.97). A nearly exponential  mass growth  has also been obtained at redshift up to 3 by,  e.g.,
 Gr{\"u}tzbauch et al. (2011)   and beyond, e.g.  Labb{\'e} et al. (2010).
 The specific star formation rate (SFR/$M_*$) (Figure \ref{scale}(c)) is thus unsurprisingly flat,
 or marginally decreasing with increasing mass.
 
 In none of Figures \ref{scale} there is compelling evidence that the relations shown have a significant
 change of slope in the considered interval of stellar mass, in contrast to the one 
 found by Huang  et al. (2012) near log$M_*\sim 9$ M$_\odot$.
 This is probably because the Local Supercluster lacks high mass galaxies due to the small sampled volume
 compared to $\alpha$.40. However, a change of slope 
 
  Good agreement is found between H$\alpha3$ and HRS (Cortese et al. 2011) by comparing our deficient (green dots) or 
 strongly deficient (red dots) objects
 with the $Virgo$ HRS subsample (large red squares in Figure \ref{scale}(a)).
 Altogether,  there is full agreement with previous surveys that on the common mass range there is a primary scaling relation between
 HI and the stellar mass, but H$\alpha3$ extends the studied range of mass to one order of magnitude fainter.   
  
 High-deficiency galaxies (see Figure \ref{scale}(a))   
 are found exclusively among the most massive systems and
 lie dangerously near to the ALFALFA detection limit already at the distance of Virgo 
 (this lack of sensitivity is at the origin of the LTGs undetected by ALFALFA considered in Section \ref{results}).
 More highly HI-deficient galaxies exist at low stellar mass, 
 but they require deeper HI observations than ALFALFA to be detected. 
 Indeed when the optically selected galaxies (see Section \ref{results}) are considered 
 and their deeper available HI observations are used to complement the ALFALFA data in
 Figure \ref{scale}(a), several objects with severe HI depletion are detected 
 at all stellar masses, including objects as faint as $10^7$ $M_{\odot}$.  
 This is consistent with the idea that hydrogen depletion occurs more easily in 
 dwarf galaxies within a cluster like Virgo (Boselli et al. 2008). 
 The time scale for gas depletion is
 short enough ($\sim$ 100 Myr) to quickly transform them into anemics that become
 gas-free, passive dEs. This is precisely
 the mechanism invoked by Boselli et al. (2008) and by Gavazzi et al. (2010) to migrate dwarf 
 star-forming systems into dEs in the neighborhood of rich galaxy clusters.
 \begin{table}[h!]
\caption{Linear regression coefficients of the relations in Figures \ref{scale}  
obtained using the bisector method (mean coefficients 
of the direct and the inverse relation).
}
\begin{center}
\begin{tabular}{lllr}
\hline
Linear regression  & $r$  & $Def_{HI}$ & Fig.  \\
\hline
  log$M_{HI}/M_*=-0.63\times$ log$M_*+5.53$ &-0.90 & normal   &(a)\\     
  log$M_{HI}/M_*=-0.71\times$ log$M_*+5.73$ &-0.95 & intermed &(a)\\
  log$M_{HI}/M_*=-0.79\times$ log$M_*+5.93$ &-0.90 & high   &(a)\\
  log$SFR=0.89\times$ log$M_*-8.92$ &0.88 & normal  &(b)\\ 
  log$SFR=0.84\times$ log$M_* -8.68$ &0.89 & intermed &(b)\\
  log$SFR=0.91\times$ log$M_*-9.86$ &0.79 & high  &(b)\\
  log$SSFR=-0.56\times$ log$M_*-5.09$ &-0.46 & normal   & (c)\\  
  log$SSFR=-0.52\times$ log$M_*-5.57$ &-0.55 & intermed  & (c)\\ 
  log$SSFR=-0.70\times$ log$M_*-3.93$ &-0.45 & high    & (c)\\ 
  log$SSFR=0.40\times$ log$M_{HI}/M_*-9.86$& 0.53 & normal   & \ref{paura} (L)\\
  log$SSFR=0.41\times$ log$M_{HI}/M_*-9.86$& 0.62 & intermed  & \ref{paura} (L)\\
  log$SSFR=0.52\times$ log$M_{HI}/M_*-9.82$& 0.55 & high      & \ref{paura} (L)\\
\hline
\hline
\end{tabular}
\end{center}
\label{fit}
\end{table}

 In our sample the specific star formation is found to depend only mildly on
 the stellar mass (Figure \ref{scale}(d), compared with Huang et al (2012) or with optical selected samples 
 (Gavazzi et al. 2002b, Schiminovich et al. 2010).
 
 The most remarkable feature of the specific star formation vs stellar mass relation is 
 the enormous scatter,  as noted by other authors (Lee at al. 2007, Bothwell et al. 2009),  
 particularly evident at low mass, where the scatter reaches 2 orders of magnitude. 
 The faintest dwarfs are amorphous objects composed of few HII regions. 
 Some of these are outliers with respect to the trends in Figures \ref{scale}, with a high SFR for a given HI mass and $M_*$.
 It is debated (Boselli et al. 2009, Fumagalli et al. 2011,  Weisz et al. 2011) 
 whether sporadic bursts of SFR occurring in the individual HII regions 
 in a stochastic manner could produce 100 \% variations in the integrated SSFR, hence
 a spread similar to the one observed in Figure \ref{scale}(c).
  The same relation was also studied  by Bothwell et al. (2009) and by Schiminovich et al. (2010) (see their Fig. 8)
 in the (optically selected) GASS survey, limited to galaxies with $10^{10}<M_*<10^{11.5}$ M$_\odot$. 
 In this mass range our data agree with their $high~SFE$ objects (see the blue line in Figure 
 \ref{scale}(c)), but in addition to these, Bothwell et al. (2009) and Schiminovich et al. (2010) 
 have many  $low ~SFE$ objects at  log SSFR $\sim$ -12 $\rm yr^{-1}$(see the red line in 
 Figure \ref{scale}(c)) that
 are simply not sampled in H$\alpha3$, which makes the overall relation steeper than in $H\alpha3$. 
 Boselli et al. (2001) also found a correlation steeper than in Figure \ref{scale}(c).

Summarizing, our analysis  has highlighted in agreement with previous studies
(e.g., Vulcani et al. 2010, Sobral et al. 2011)
that the stellar mass is a fundamental parameter in driving the observed scaling relations. The
effect of the environment is then superimposed on this scaling relations and is more evident 
in the relations that involve the mass of neutral hydrogen, which more easily prone to the effects
of the environment when quantified using the HI deficiency.

\section{Overview of the SFR in the Local Volume}
\label{result}

\begin{figure}[h!]
\centering
\includegraphics[width=9cm]{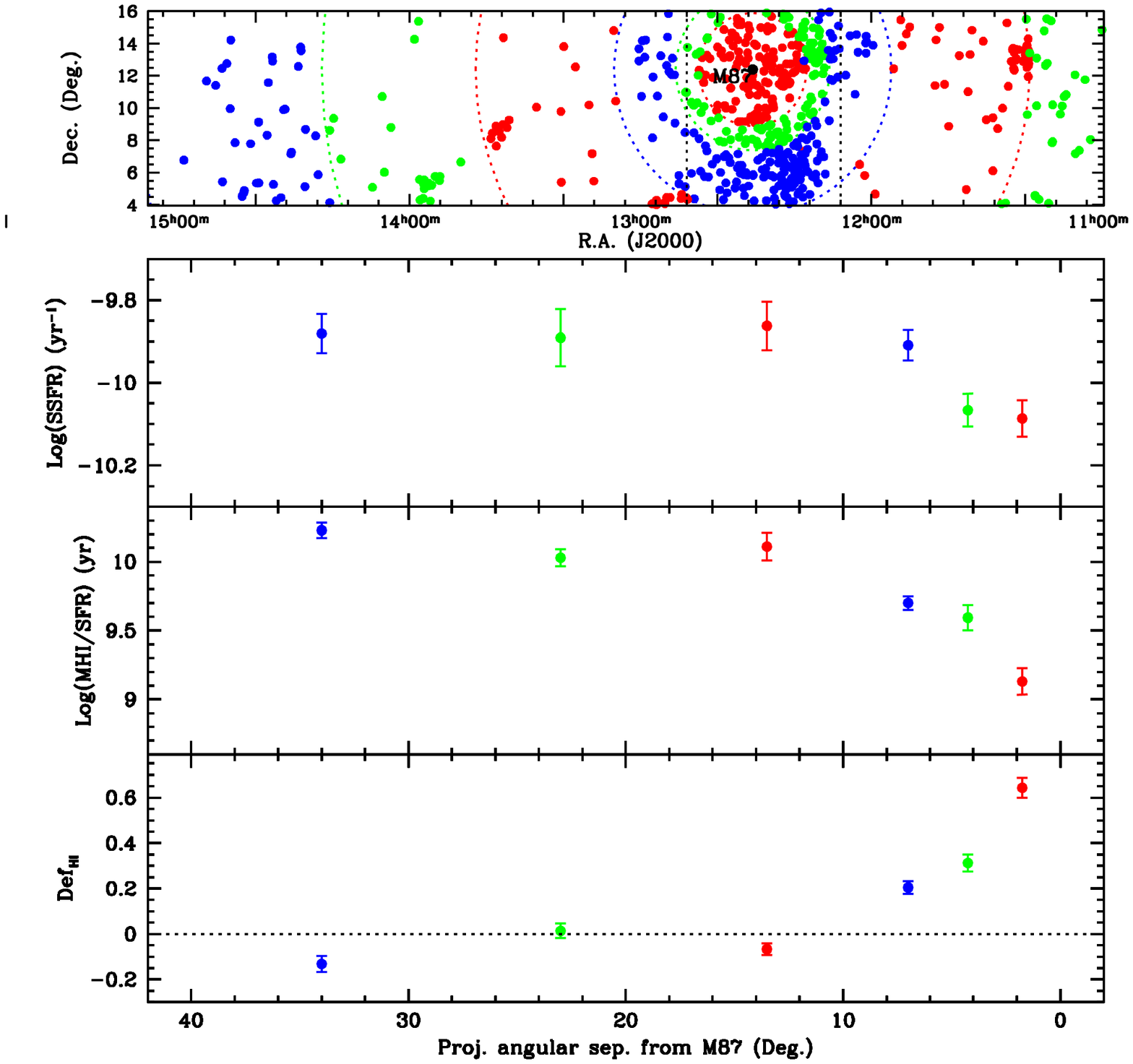}
\includegraphics[width=9cm]{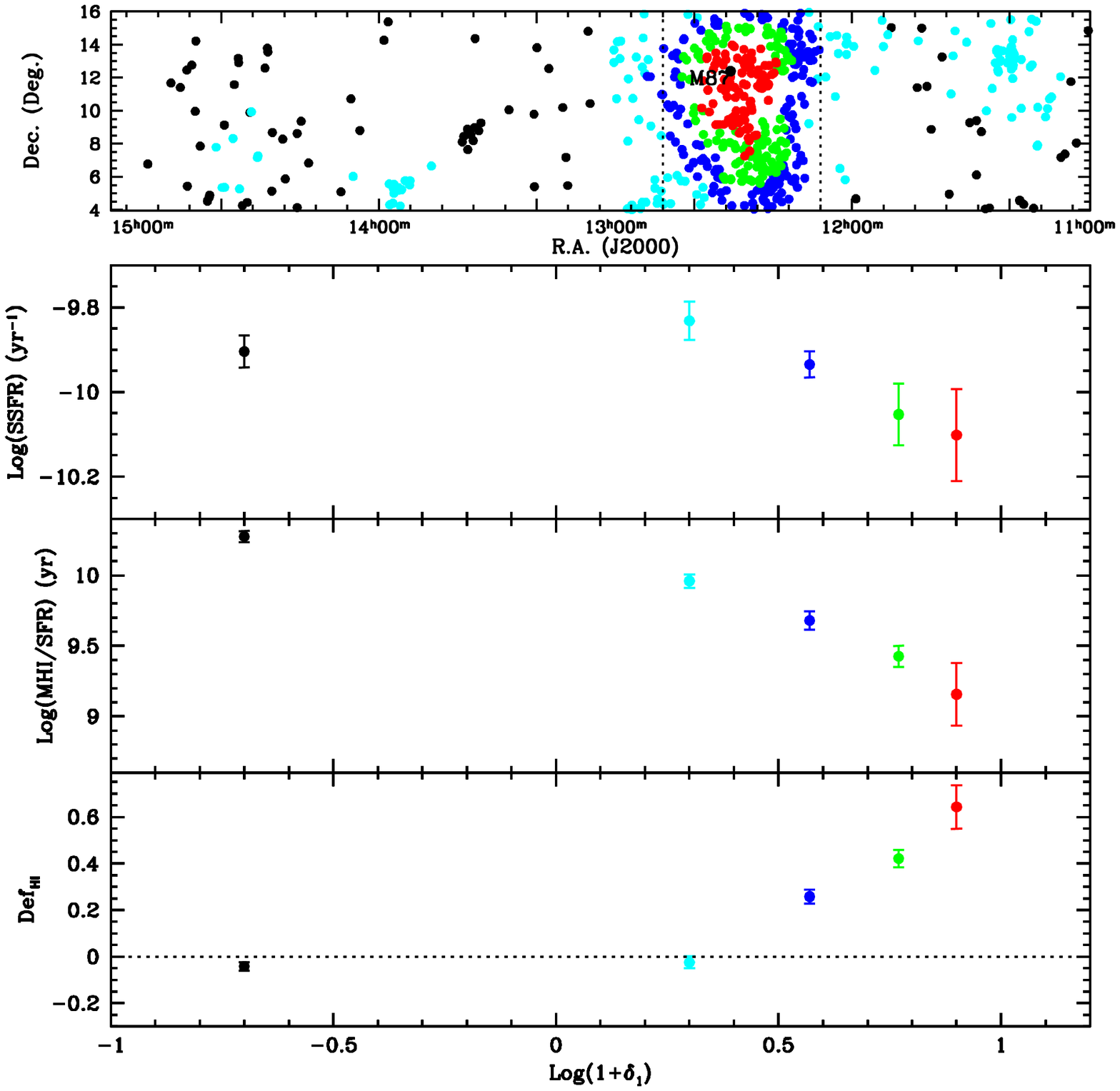}
 \caption{Top panel: the sky distribution of LTGs within seven annuli of increasing radius from M87. 
 Galaxies in each ring are given with a different color. (b): SSFR; 
 (c): SFR per unit HI mass; (d) HI-deficiency  as function of the projected 
 angular separation from M87. Bottom panel: same as above in bins of local galaxy density. 
 This analysis is limited to galaxies with projected distance $<45^o$ from M87, i.e,
 20 objects in the interval 15:00:00$\rm <RA<$16:00:00) were excluded because they
 do not constitute a statistically meaningful subset.
 }
\label{global}
\end{figure}

We now illustrate the global role of the Virgo 
cluster on the star formation properties of the Local Supercluster as a whole, 
as shown in Figure \ref{global}. Only the population of LTGs is considered, 
meaning we disregard ETGs that would obviously trace the
phenomenon of morphology segregation in the Virgo cluster. 
The top panel shows the celestial distribution of LTGs in the present work.  
Annuli centered on M87 are drawn with dashed lines, chosen to contain approximately an equal number of galaxies\footnote{Figure \ref{global} 
was obtained adding to the ALFALFA-detected LTGs
the undetected LTGs that were optically selected with HI and H$\alpha$ measurements taken
from GOLDMine.}. 
After discarding galaxies that do not properly belong to cluster A or B (background groups, such as clouds M, W),
within each annulus we compute 
the median specific SFR (SFR per unit stellar mass) (second panel).
The depletion or Robert's time (HI mass divided by SFR) shown in the third panel 
gives the duration of the remaining star formation activity at the 
present rate for the available HI content.
The inverse of the latter quantity is sometime called the star formation efficiency.
The HI deficiency is given in the bottom panels. These quantities are plotted in bins of projected distance
from M87 (top panels) and in bins of local galaxy density (bottom panels). 

There is a significant radial dependence of the HI content (the strongest of all),  
of the specific star formation rate, and therefore of the depletion time.
LTGs in the Local Supercluster located up to 20 degrees away from Virgo have a life expectation
longer than $10^{10}$ yr (i.e, almost one Hubble time) if they continue to burn their hydrogen at the present rate,
while galaxies near to M87 will remain active $10^{9.3}$ yr (less than 2 Gyr). This derives from the reduced 
gas reservoir of galaxies in the cluster environment (HI-deficiency) and appears to be 
associated with a significant suppression of the star formation rate. 

\section{Discussion}
\label{discussion}

 The scaling relations between the gaseous (HI) mass and the star formation rate as a function of 
 the stellar mass,  in intervals of HI deficiency 
 are analyzed in Figure \ref{scale} and are found to be consistent with previous 
 determinations based on HI and on optically selected samples.

\begin{figure*}
   \centering
   \includegraphics[width=8cm,height=8cm]{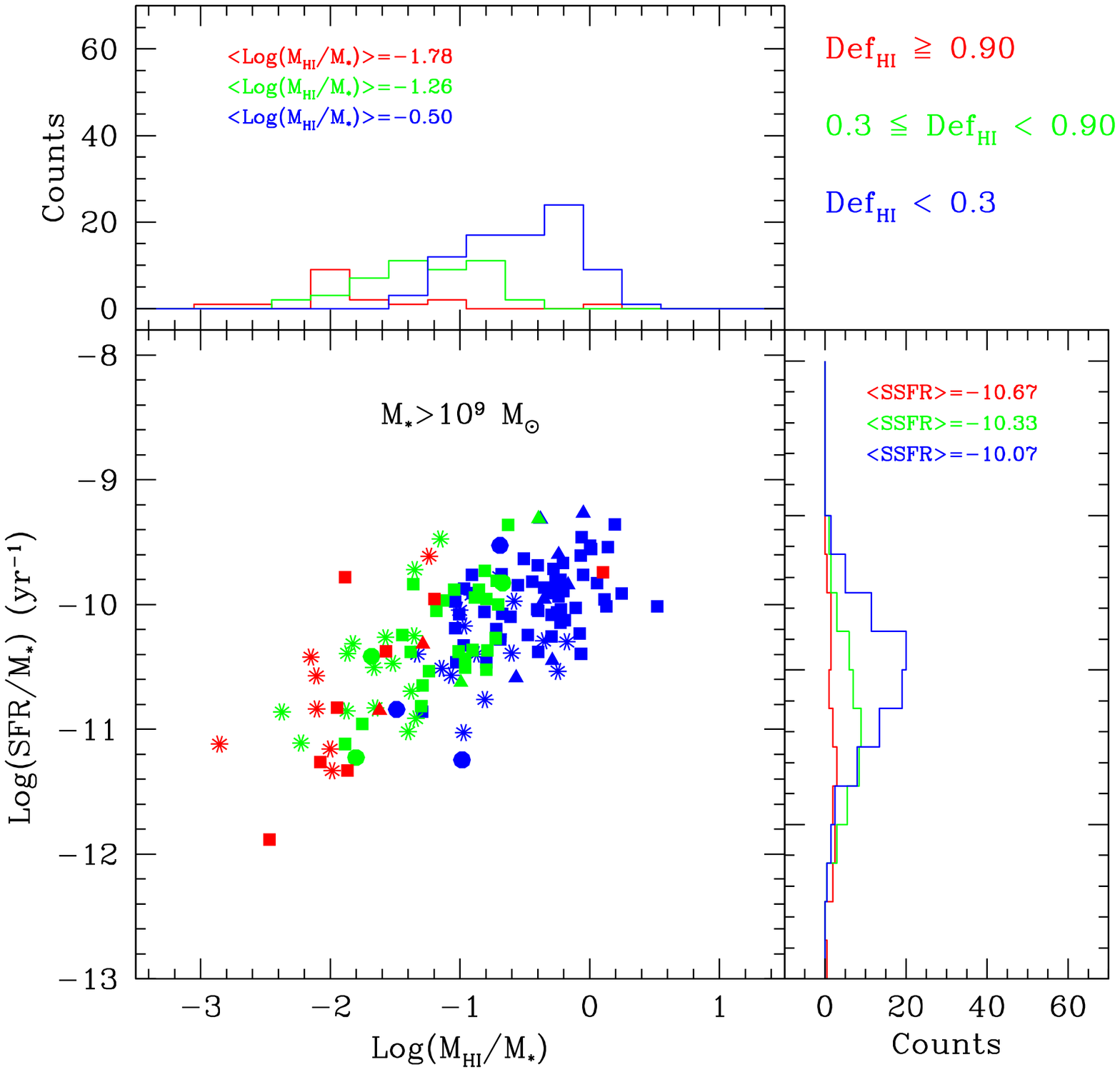}
   \includegraphics[width=8cm,height=8cm]{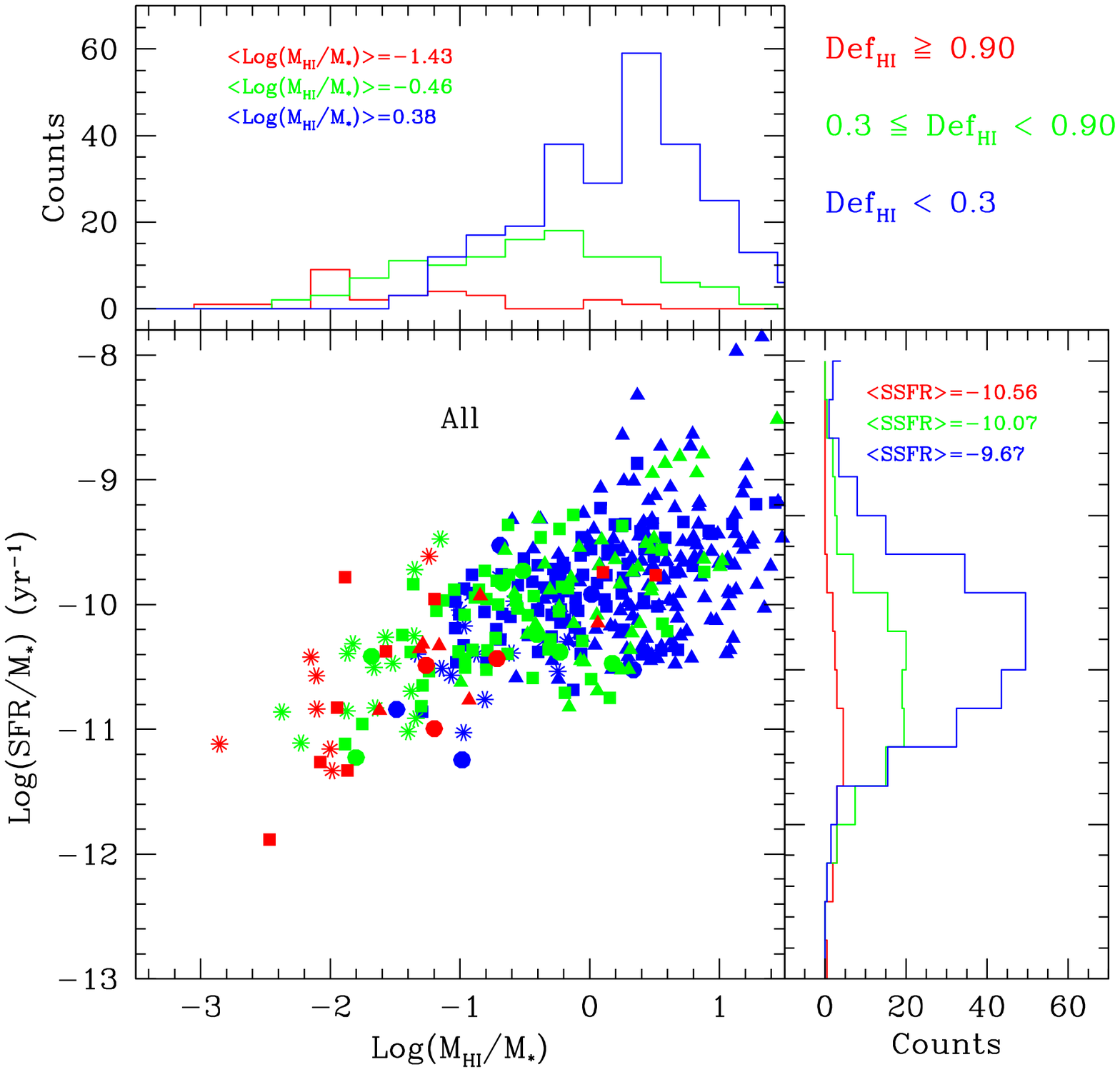}
   \caption{(Left panel) Relation between the specific star formation rate and the gas fraction in three 
     classes of increasing $Def_{HI}$, excluding objects fainter than $M_*=10^9$ $M_\odot$. (Right panel) Same  
     as in the left panel, but including fainter objects. 
     The histograms drawn on top and to the right show the counts of galaxies with given HI  gas fraction
     and SSFR, with the mean values given in three bins of $Def_{HI}$.}
   \label{paura}
 \end{figure*}

 The environmental conditions occurring in a young, still evolving cluster of galaxies such as the 
 Virgo cluster have profound consequences  on the morphological composition of its member galaxies,
 producing the well known phenomenon named morphology segregation (Dressel 1980). 
 Moreover,  it induces observable modifications to the population of LTGs, even when they are considered alone:
 their HI content and to a lesser extent their specific star formation are significantly reduced compared
 to LTGs found in relative isolation. 
 To quantify the quenching introduced by the environment on the gas content and on the SFR independently of the stellar mass,
we plot in the central panel of
Figure \ref{paura} (a,b) the relation of the specific star formation versus gas fraction.
This Figure was obtained by dividing both the SFR and the HI mass 
by the stellar mass, so that both axes contain normalized quantities.  
In other words, this figure contains the residual correlation excluding 
the scaling law with stellar mass.
The symbols are color-coded according to the $Def_{HI}$ parameter.
The histograms in the side panels enable to estimate the average gas fraction and SSFR in
three bins of HI deficiency.  
While  the left panel contains only giant galaxies (log$M_*>9$ M$_\odot$), the right panel 
shows all galaxies. Both panels show that even after subtracting the scaling law with stellar mass
, a significant residual (second-order) dependence of the SSFR on the HI-deficiency
remains: gas-rich objects that contain as much mass in HI and stars (log $M_{HI}/M_*$=0) have SSFR ten times
higher than galaxies that contain 100 times less HI relative to stars. 
The gas-rich, high SSFR systems are exclusively found among unperturbed galaxies ($Def_{HI}\leq 0.3$),
while gas-poor objects with a lower SSFR are exclusively highly perturbed ($Def_{HI}\geq 0.9$) members 
of Virgo. The comparison between the two panels emphasizes the large scatter in the SSFR of low mass objects
mentioned earlier. These faint systems preferentially have a high gas fraction, normal or 
intermediate HI-deficiency and an erratic specific star formation rate over 2 dex,
in contrast to high stellar mass
 systems, whose specific star formation correlates better with the gas fraction (see Boselli et al. 2009).
  In summary,  once the scaling relation with stellar mass is neutralized, 
 it is found  that  by decreasing the gas fraction 
 by 2 dex, the specific star formation rate decreases by one dex, i.e, 
 the effect of the environment is ten times as effective at removing the HI gas than at quenching 
 the star formation.
  
 This is consistent with the finding of Figure \ref{global}, which shows 
 that the star formation efficiency ($M_{HI}$/SFR) decreases going toward the Virgo cluster,
 i.e, $M_{HI}$ decrease more than SFR, implying that the cluster produces 
 a significant perturbation to the Local Supercluster, even on its member late-type galaxies,
 particularly efficient at reducing their HI content and to a lesser extent
 at quenching their present star formation rate.
 
 In Section \ref{results} we showed that there is evidence of a continuous sequence 
 of transformations that make LTGs gradually gas-poorer and redder in their disk components
 while they approach the Virgo cluster. Meanwhile, their star formation
 shrinks to the circumnuclear or even nuclear scale.
 The last feature calls for a truncation process of the star formation that is
 inconsistent with the so-called strangulation (Boselli et al. 2006, 2008) and consistent with ram pressure.
  
 HI ablation occurs on very brief time scales of 100-200 Myr for low-mass galaxies and 
 300 Myr for massive galaxies. 
 An example is NGC 4569, analyzed by Boselli et al. (2006). 
 In spite of being very HI-poor (85\% of the HI gas has been removed from this galaxy by ram-pressure), 
 neverless it is only marginally $\rm H_2$ deficient (Fumagalli et al. 2009) and its
 global star formation is only reduced by a factor of two. This is consistent with the 
 fact that most of the gas has been removed from the outer disk, 
 while in the inner regions the star formation continues sustained by copious $\rm H_2$, 
 coexistent with an inner HI shell.
 The time-scales for ram-pressure stripping in this massive galaxy is 
 consistently estimated to be $\sim$ 300 Myr by 
 Vollmer et al. (2004) and between 100 and 400 Myr by Boselli et al. (2006).
 Estimates of 200 Myr were derived for two dwarf galaxies in the Virgo cluster: VCC 1217 (Fumagalli et al. 2011b)
 and VCC 1249 in the M49 group (Arrigoni Battaia et al. 2012). 
 
 The ongoing infall rate of galaxies on the Virgo cluster can be
 obtained by dividing the number of HI rich LTGs that still exist in the Virgo cluster 
 by the time scale of the HI stripping mechanism. Table \ref{infall} contains two estimates:
 for $>10^7$ $M_\odot$ and for $>10^9$ $M_\odot$ (the second interval is to compare our estimate with a similar estimate obtained for the Coma cluster in Paper III).
 In the two cases we obtain a current infall rate of 400 - 100 galaxy $\rm Gyr^{-1}$ in the two mass bins 
 (the former consistent with  300 galaxy $\rm Gyr^{-1}$ found by Boselli et al. (2008) for Virgo and the latter consistent with
 the estimate for Coma).
 
 An alternative method for estimating the infall rate can be achieved using the number of LTGs that are currently in the HI-poor phase 
 divided by the time scale for  complete hydrogen exhaust: 0.5 Gyr for low-mass galaxies 
 (Boselli et al. 2008) and somewhat longer for massive systems. 
 This time scale is estimated assuming that a complete HI and $\rm H_2$ ablation will 
 occur during the second passage through the cluster center.
 This is considerably shorter than $\sim 2\pm 1$ Gyr (Bigiel et al. 2008): 
 the time scale for complete consumption of the $\rm H_2$  
 due to star formation in intermediate HI deficiency galaxies, i.e,  
 without an extended reservoir of atomic hydrogen that can replenish the inner 
 HI fuel consumed by the star formation (Fumagalli et al. 2009).
 The derived infall rates adopting these figures are consistent with the previous estimates.
 
 Finally, assuming that all galaxies (LTGs+ETGs) were accreted by the Virgo cluster at a rate that remained approximately constant
 in time,  and that massive galaxies require an additional 1-2 Gyr after a stripping event
 to migrate from the blue cloud to the transition region and to the red sequence 
 (Boselli et al. 2008; Cortese \& Hughes  2009; Gavazzi et al. 2010),  we infer that the age of the Virgo cluster is 
 roughly 2 Gyr, implying a cluster formation at $z \sim$ 0.2, much later than $z \sim$ 1 derived using similar arguments for the Coma cluster
 in Paper III. 
 Of course, these are just order-of-magnitude estimates of the accretion rate on the Virgo cluster. For example,
 cluster B (M49) will soon merge with cluster A (M87) to form an even more ralaxed and rich Virgo cluster,
 contributing with many galaxies that have been pre-processed in the group itself.   
 The significant age difference found between Virgo and Coma is consistent with the high contrast in the observed
 X-ray properties of the two clusters (Forman \& Jones 1982), suggesting a marked difference in their evolutionary stages.   
    \begin{table}[h!]
    \caption{Build-up of the Virgo cluster from various estimators. The first number refers to objects
     with stellar mass $>10^7$ $M_\odot$, the second to those with $>10^9$ $M_\odot$ (comparable with Coma)
    }
    \begin{center}
    \begin{tabular}{llll}
    \hline
                            & All                     	      & HI-poor 			& HI-rich	      \\
    \hline
    N ($>10^7$ $M_\odot$)                       & 738                       & 160 		& 93			    \\
    t (Gyr)                  & 2 $\pm$ 1           & 0.5 $\pm$ 0.4	& 0.2 $\pm$ 0.1      \\
    Infall rate (Gyr$^{-1}$) & 369 $\pm$ 184    & 320 $\pm$ 190 	& 465 $\pm$ 230            \\
    \hline
    N ($>10^9$ $M_\odot$)                       & 210                      &  73			&  36			    \\
    t (Gyr)                  &  2 $\pm$ 1          &  0.7 $\pm$ 0.4	&  0.3 $\pm$ 0.1     \\
    Infall rate (Gyr$^{-1}$) &  105 $\pm$ 53   &  104 $\pm$ 61	&  120 $\pm$ 45      \\
    \hline
    \hline
    \end{tabular}
    \end{center}
    \label{infall}
    \end{table}

\section{Summary and conclusions}
\label{concl}

 The wealth of data provided by ALFALFA in the stretch of the Local Supercluster  visible in spring, which contains 
 the Virgo cluster,  which were
 followed-up with H$\alpha$ imaging observations, enabled us to analyze the relations 
 between the HI  gas, the stellar content and the 
 rate at which the gas is currently fueling the formation of new stars in spiral galaxies. \\ 
 Given the sensitivity of ALFALFA and of H$\alpha3$ at the distance of Virgo, our analysis comprises 
 galaxies with a stellar mass in the range $10^{7.5}<M_*<10^{11.5}$ M$_\odot$, a gas mass 
 $10^{7.7}<M_{HI}<10^{10}$ M$_\odot$ and a star formation rate in excess of $10^{-3}$ M$_\odot \rm yr^{-1}$.
 The main results of the present investigation can be summarized as follows:
 \begin{itemize}
\item {In addition to morphology segregation, the Virgo cluster represents a significant perturbation on the Local Supercluster 
for the gas content  and the star formation properties of its member galaxies of late-type, such that on average
the HI content is reduced by a factor of four and the specific star formation rate by a factor of two.
}
  \item {LTGs in the Local Supercluster have progressively redder color with decreasing 
  HI gas content, from normal deficient galaxies to  objects that are undetected by ALFALFA. 
  The gas-ablation mechanism also governs the star formation properties. 
 }
   \item {Highly HI deficient and undetected LTGs have ongoing nuclear
   and circumnuclear (truncated) star formation, while HI-rich LTGs show star formation extended on the disk scale. 
   We propose that the quenching mechanism responsible for these features is ram pressure.
    }
 \item {
 The fundamental scaling relation (studied in Section \ref{scale}) between the HI mass and the stellar mass 
 in normal spiral galaxies is such that  $M_{HI}\sim M_*^{0.5}$. This implies that 
 increasingly more massive galaxies have progressively lower HI gas fraction, consistent with the idea that they
 have transformed most of their gas into stars at higher $z$, in agreement  with the $downsizing$ scenario. 
 }
  \item {The scaling relation between $M_{HI}$ and $M_*$ followed by galaxies in the Virgo cluster 
  is significantly offset toward lower gas content, compared to that of HI-normal galaxies.
  }
 \item {Once the scaling law with stellar mass is removed by analyzing the correlation between the specific star 
  formation and the 
 gas fraction, it is found that a residual second-order correlation exists between these quantities. 
 The effect of the environment is
 ten times stronger at producing the gas depletion than at quenching the star formation.
 }
 \item {The present infall rate of $\sim$ 400 galaxies per Gyr (of $\geq 10^7$ M$_\odot$ objects), 
 inferred by the number of healthy spirals present in the Virgo cluster,
 is consistent with the number of galaxies of all types (of similar mass) if the infall process has
 been acting for approximately 2 Gyr, about three times shorter than on the Coma cluster (Paper III).
 This is consistent with the idea that Virgo is a young cluster caught in an early evolutionary stage.
 }
 \end{itemize}
 
\begin{acknowledgements}
We thank Luca Cortese and Barbara Catinella
for their comments on an early version of the manuscript, and
Massimo Dotti and Emanuele Ripamonti for useful discussions.
We also thank Silvia Fabello for her contribution to the observations and Shan Huang 
for providing original data.
The authors would like to acknowledge the work of the entire ALFALFA collaboration team
in observing, flagging, and extracting the catalog of galaxies used in this work.
This research has made use of the GOLDMine database (Gavazzi et al. 2003)
and of the NASA/IPAC Extragalactic Database (NED) which is operated 
by the Jet Propulsion Laboratory, California Institute of Technology, under contract with the
National Aeronautics and Space Administration. 
We wish to thank the anonymous referee, whose criticism helped us to improve the manuscript.
Funding for the Sloan Digital Sky Survey (SDSS) and SDSS-II has been provided by the 
Alfred P. Sloan Foundation, the Participating Institutions, the National Science Foundation, 
the U.S. Department of Energy, the National Aeronautics and Space Administration, 
the Japanese Monbukagakusho, and 
the Max Planck Society, and the Higher Education Funding Council for England. 
The SDSS Web site is \emph{http://www.sdss.org/}.
The SDSS is managed by the Astrophysical Research Consortium (ARC) for the Participating Institutions. 
The Participating Institutions are the American Museum of Natural History, Astrophysical Institute Potsdam, 
University of Basel, University of Cambridge, Case Western Reserve University, The University of Chicago, 
Drexel University, Fermilab, the Institute for Advanced Study, the Japan Participation Group, 
The Johns Hopkins University, the Joint Institute for Nuclear Astrophysics, the Kavli Institute for 
Particle Astrophysics and Cosmology, the Korean Scientist Group, the Chinese Academy of Sciences (LAMOST), 
Los Alamos National Laboratory, the Max-Planck-Institute for Astronomy (MPIA), the Max-Planck-Institute 
for Astrophysics (MPA), New Mexico State University, Ohio State University, University of Pittsburgh, 
University of Portsmouth, Princeton University, the United States Naval Observatory, and the University 
of Washington.\\
G. G. acknowledges financial support from the italian MIUR PRIN contract 200854ECE5.
R.G. and M.P.H. are supported by US NSF grants AST-0607007 and AST-1107390
and by a Brinson Foundation grant.
Support for M.Fumagalli was provided by NASA through Hubble Fellowship grant HF-51305.01-A
awarded by the Space Telescope Science Insititute, which is operated by the Association of Universities
for Research in Astronomy, Inc., for NASA, under contract NAS 5-26555.
\end{acknowledgements}

\end{document}